%% file: main.tex
\documentclass[acmsmall,screen]{acmart}
\usepackage[utf8]{inputenc}
\usepackage{graphicx}
\usepackage{amsmath,amssymb}
\usepackage{mathtools}
\usepackage{relsize}
\usepackage{stmaryrd}
\usepackage{rotating}
\usepackage{paralist}
\usepackage{epstopdf}
\usepackage{algpseudocode}
\algnotext{EndFor}
\algnotext{EndIf}
\algnotext{EndProcedure}
\usepackage{algorithm}
\usepackage{enumitem}
\usepackage{caption}
\usepackage{tikz}
\DeclareCaptionType{copyrightbox}
\usepackage{subcaption}
\usepackage{lmodern}
\usepackage{graphicx}
\usepackage{appendix}
\usepackage{soul}
\usepackage{tabularx}
\usepackage{booktabs}
\usepackage{mathrsfs}
\usepackage{amsbsy}
\usepackage{xspace}
\usepackage{harpoon}

\usepackage{lstlinebgrd}
\usepackage{listings}
\usepackage{capt-of}
\usepackage{pst-node}
\usepackage{tikz-cd} 
\usepackage{slashbox}

\makeatletter\if@ACM@journal\makeatother
\acmYear{2018}
\acmMonth{4}
\startPage{1}
\else\makeatother
\acmYear{2018}
\acmISBN{978-x-xxxx-xxxx-x/YY/MM}
\acmDOI{10.1145/nnnnnnn.nnnnnnn}
\startPage{1}
\fi

\setcopyright{none} 

\setlist[itemize]{leftmargin=*,noitemsep, topsep=0.5pt}
\setlist[enumerate]{leftmargin=*,noitemsep, topsep=0.5pt}

\usepackage{ascii} 
\usepackage[T1]{fontenc}

\usepackage{booktabs}
\usepackage{balance}

\usepackage{defs}
\usepackage{mathpartir}
\usetikzlibrary{arrows}
\usetikzlibrary{shapes}
\usetikzlibrary{matrix}

\input{macros}

\newcommand\xqed[1]{%
  \leavevmode\unskip\penalty9999 \hbox{}\nobreak\hfill
  \quad\hbox{#1}}
\newcommand\demo{\xqed{$\triangle$}}

\renewcommand\footnotetextcopyrightpermission[1]{}

\begin{document}

\title{Efficient Differentiable Programming in a Functional Array-Processing Language}

\author{Amir Shaikhha}
\affiliation{
  \institution{EPFL}            
  \city{Lausanne}
  \country{Switzerland}
}
\email{amir.shaikhha@epfl.ch}          

\author{Andrew Fitzgibbon}

\affiliation{
  \institution{Microsoft HoloLens}
  \city{Cambridge}
  \country{United Kingdom}
}
\email{awf@microsoft.com}

\author{Dimitrios Vytiniotis}

\affiliation{
  \institution{Microsoft Research}
  \city{Cambridge}
  \country{United Kingdom}
}
\email{dimitris@microsoft.com}

\author{Simon Peyton Jones}

\affiliation{
  \institution{Microsoft Research}
  \city{Cambridge}
  \country{United Kingdom}
}
\email{simonpj@microsoft.com}

\author{Christoph Koch}
\affiliation{
  \institution{EPFL}            
  \city{Lausanne}
  \country{Switzerland}
}
\email{christoph.koch@epfl.ch}

\renewcommand{\shortauthors}{A. Shaikhha et al.}

\begin{abstract}

We present a system for the automatic differentiation of a higher-order functional array-processing language. The core functional language underlying this system simultaneously supports both source-to-source automatic differentiation and global optimizations such as loop transformations. Thanks to this feature, we demonstrate how for some real-world machine learning and computer vision benchmarks, the system outperforms the state-of-the-art automatic differentiation tools.
\end{abstract}
\maketitle
\input{intro}
\input{overview}
\input{fsmooth_d}

\input{efficient}

\input{exp}
\input{related}
\input{conclusion}

\bibliographystyle{ACM-Reference-Format}
\bibliography{ref} 






\end{document}

%% file: macros.tex
\lstdefinelanguage{Scala}%
{morekeywords={abstract,%
  case,catch,char,class,%
  def,else,extends,final,finally,for,%
  if,import,implicit,%
  match,module,%
  new,null,%
  object,override,%
  package,private,protected,public,%
  for,public,return,super,%
  this,throw,trait,try,type,%
  val,var,%
  with,while,%
  yield%
  },%
  sensitive,%
  morecomment=[l]//,%
  morecomment=[s]{/*}{*/},%
  morestring=[b]",%
  morestring=[b]',%
  showstringspaces=false%
}[keywords,comments,strings]%

\definecolor{listingbg}{RGB}{240, 240, 240}

\newcommand{\lett}{\cod{let}}
\newcommand{\inn}{\cod{in}}
\newcommand{\iif}{\cod{if}}
\newcommand{\then}{\cod{then}}
\newcommand{\elsee}{\cod{else}}

\newcommand{\tab}{\;\;\;}

\newcommand{\lafsharp}{$\widetilde{\textsc{F}}$}
\newcommand{\fsmooth}{\lafsharp}

\newcommand{\ladsl}{$\widetilde{\textsc{M}}$}
\newcommand{\system}{$\text{d}\widetilde{\textsc{f}}$\xspace}

\newcommand{\viteratek}{\cod{ifold}}
\newcommand{\vifoldk}{\viteratek}

\newcommand{\vbuildk}{\cod{build}}
\newcommand{\vlengthk}{\cod{length}}

\newcommand{\vpairk}{\cod{pair}}
\newcommand{\vfstk}{\cod{fst}}
\newcommand{\vsndk}{\cod{snd}}

\newcommand{\expr}{\text{e}}

\newcommand{\exprind}[1]{\text{$\text{\expr}_{#1}$}}

\newcommand{\vbuild}[2]{\text{\vbuildk{} #1 #2}}
\newcommand{\vget}[2]{#1[#2]}
\newcommand{\vlength}[1]{\vlengthk{}\,#1}
\newcommand{\vifthenelse}[3]{\text{\iif{} #1 \then{} #2 \elsee{} #3}}

\newcommand{\viterate}[3]{\text{\viteratek{} #1 #2 #3}}
\newcommand{\vabsk}{\cod{fun}}
\newcommand{\vabs}[2]{\text{\vabsk{} #1 \cod{->} #2}}

\newcommand{\vapp}[2]{\text{#1 #2}}
\newcommand{\vlet}[3]{\text{\lett{} #1 = #2 \cod{in} #3}}

\newcommand{\vpair}[2]{\text{(#1, #2)}}

\newcommand{\vconst}[1]{\text{\cod{#1}}}

\newcommand{\typewrapper}[1]{\text{\cod{#1}}}
\newcommand{\typewrappernonterm}[1]{\textnormal{#1}}
\newcommand{\typemat}{\typewrappernonterm{M}}

\newcommand{\typepair}[2]{\text{#1 $\times$ #2}}
\newcommand{\typet}{\typewrappernonterm{T}}

\newcommand{\typeind}[1]{\text{$\typet{}_{#1}$}}

\newcommand{\funarrow}{\text{$\Rightarrow$}}

\newcommand{\typefunone}[2]{\text{#1 $\Rightarrow$ #2}}
\newcommand{\typeindex}{\typewrapper{Index}}
\newcommand{\typecard}{\typewrapper{Card}}
\newcommand{\typedouble}{\typewrapper{Double}}

\newcommand{\typebool}{\typewrapper{Bool}}
\newcommand{\typearray}[1]{\typewrapper{Array<#1>}}
\newcommand{\typevector}{\typewrapper{Vector}}
\newcommand{\typematrix}{\typewrapper{Matrix}}

\def\typenum{\typewrappernonterm{Num}}
\newcommand{\typedoubled}{\typewrapper{DoubleD}}
\newcommand{\typevectord}{\typewrapper{VectorD}}
\newcommand{\typematrixd}{\typewrapper{MatrixD}}

\newcommand{\evalsto}{\text{ $\leadsto$ }}

\newcommand{\valcard}{\text{N}}

\def\tabt{\hspace*{0.35cm}}

\def\vcaddcard{\text{$+$}}
\def\vcsubcard{\text{$-$}}

\def\vcget{\text{\cod{get}}}
\def\vclength{\text{\cod{length}}}

\newcommand{\difftranstwo}[2]{\text{$\mathcal{D}\llbracket$#1$\rrbracket$}}
\newcommand{\difftrans}[1]{\difftranstwo{#1}{dx}}
\newcommand{\difftransp}[1]{\text{$\mathcal{V}\llbracket$#1$\rrbracket$}}
\newcommand{\difftranst}[1]{\text{$\mathcal{T}\llbracket$#1$\rrbracket$}}
\newcommand{\difftranstype}[1]{\text{$\mathcal{D_{\text{T}}}\llbracket$#1$\rrbracket$}}

\newcommand{\diffvarprefix}[1]{\text{$\overrightharp{\text{#1}}$}}

\newcommand{\pterm}[1]{\text{\vfstk{} (#1)}}
\newcommand{\dterm}[1]{\text{\vsndk{} (#1)}}

\newcommand{\adpair}[2]{\text{(#1, #2)}}
\newcommand{\diffk}{\cod{diff}}
\newcommand{\vdiffk}{\cod{vdiff}}
\newcommand{\mdiffk}{\cod{mdiff}}
\newcommand{\diff}[1]{\text{\diffk{}~#1}}
\newcommand{\gradk}{\cod{grad}}
\newcommand{\mgradk}{\cod{mgrad}}
\newcommand{\grad}[1]{\text{\gradk{}~#1}}
\newcommand{\jacobk}{\cod{jacob}}
\newcommand{\jacob}[1]{\text{\jacobk{}~#1}}
\newcommand{\derivk}{\cod{deriv}}
\newcommand{\deriv}[2]{\text{\derivk{} #1 #2}}

\newcommand{\codespace}{\vspace{0.2cm}}

\newcommand{\forwardvar}[1]{\overset{\rightharpoonup}{#1}}
\newcommand{\reversevar}[1]{\overset{\leftharpoonup}{#1}}

%% file: intro.tex
\begin{quote}
{\em
... in the summer of 1958 John McCarthy decided to investigate differentiation as an interesting symbolic computation problem, which was difficult to express in the primitive programming languages of the day. 
This investigation led him to see the importance of functional arguments and recursive functions in the field of symbolic computation. }
From Norvig~\cite[p248]{Norvig92}.
\end{quote}

\section{Introduction}
Functional programming (FP) and automatic differentiation (AD) have been natural partners for sixty years, and major functional languages all have elegant automatic differentiation packages~\cite{elliott2009beautiful,baydin2015automatic,karczmarczuk1999functional}.
With the increasing importance of numerical engineering disciplines such as machine learning, speech processing, and computer vision, there has never been a greater need for systems which mitigate the tedious and error-prone process of manual coding of derivatives.  However the popular packages (TensorFlow, CNTK) all implement clunky (E)DSLs in procedural languages such as Python and C++.  
Why?  One reason is that the FP packages are slower than their imperative counterparts, by many orders of magnitude~\cite{srajerbenchmark}, because modern applications depend heavily on array processing, with vectors, matrices, and tensors as the canonical datatypes.
In contrast, AD for FP has generally handled only scalar workloads efficiently~\cite{karczmarczuk1999functional}.

Our key contribution in this paper is to take a recently introduced F\# subset designed for efficient compilation of array-processing workloads, and to augment it with vector AD primitives, yielding a functional AD tool that is competitive with the best C/C++ and Fortran tools on many benchmarks, and considerably faster on others.

\subsection{The problem we address}


Automatic differentiation is one of the main techniques for automating the process of computing derivatives. This technique systematically applies the chain rule, and evaluates the derivatives for the primitive arithmetic operations (such as addition, multiplication, etc.). 
One of the main advantages of automatic differentiation over its main competitive technique, \textit{symbolic differentiation}, is the constant-time overhead of the differentiated program with respect to the original code.
Symbolic differentiation can lead to code explosion if one is not careful about sharing, and requires a closed-form representation of the programs~\cite{baydin2015automatic}.

There are two approaches for implementing AD. 
Forward-mode AD computes the derivative part (tangent part) alongside the original computation while making a forward pass over the program. Reverse-mode AD makes a forward pass to compute the original part of the program, followed by a backward pass for computing the derivative part (adjoint part). 
We present these two techniques through an example.

\noindent \textbf{Example.} Consider the function $f\big(x_1, x_2\big) = ln\big(x_1\big) + sin\big(x_2\big)$, for which we would like to compute the partial derivatives with respect to $x_1$ at point $x_1=1$ and $x_2=3$.
First let us name each intermediate expression
with a variable $v_i$:

\begin{tabular}{l}
$f\big(x_1, x_2\big) = $\tab \lett \,$v_1 = ln\big(x_1\big)$ \\
\tab \lett \,$v_2 = sin\big(x_2\big)$ \\
\tab \lett \,$y = v_1 + v_2$ \\
\tab $y$
\end{tabular}

\noindent This function is computed as follows:

\codespace{}
\begin{center}
\begin{tabular}{r c l c l}
$v_1$ & = & $ln\big(1\big)$ & = & $0$ \\
$v_2$ & = & $sin\big(3\big)$ & = & $0.1411$ \\
$y$ & = & $0 + 0.1411$ & = & $0.1411$ \\
\end{tabular}
\end{center}

\codespace{}
\noindent To compute the derivative of this function using the forward-mode AD, we associate the derivative $\forwardvar{v_i}=\frac{\partial v_i}{\partial x_1}+\frac{\partial v_i}{\partial x_2}$ to each variable $v_i$.
As we are computing the partial derivative of $f$ with respect to $x_1$, we have $\forwardvar{x_1} = 1$ and $\forwardvar{x_2} = 0$.
By applying the chain rule, the evaluation trace for the derivative of this function is as follows:

\codespace{}
\begin{center}
\begin{tabular}{r c l c l c l c l}
$\forwardvar{v_1}$ & = & $\forwardvar{x_1} \times \frac{\partial \, ln\big(x_1\big)}{\partial x_1}$ & = &
$\frac{\forwardvar{x_1}}{x_1}$ & = & $\frac{1}{1}$ & = & 1 \\
$\forwardvar{v_2}$ & = & $\forwardvar{x_2} \times \frac{\partial \, sin\big(x_2\big)}{\partial x_2}$ & = & 
$\forwardvar{x_2} \times cos\big(x_2\big)$ & = & $0 \times cos\big(3\big)$ & = & 0 \\
$\forwardvar{y}$ & = & $\forwardvar{v_1} \times \frac{\partial \big(v_1 + v_2\big)}{\partial v_1}+\forwardvar{v_2} \times \frac{\partial \big(v_1 + v_2\big)}{\partial v_2}$ & = &
$\forwardvar{v_1} + \forwardvar{v_2}$ & = & 1 + 0 & = & 1 \\
\end{tabular}
\end{center}

\codespace{}
\noindent 
To compute the derivative of this function using the reverse-mode AD, we associate the \textit{adjoin} term $\reversevar{v_i} = \frac{\partial y}{\partial v_i}$ to each variable $v_i$. 
As a result, if we are interested in computing the partial derivative of function $f$ with respect to $x_1$, we have to compute the value of $\reversevar{x_1}$.
To do so, we have to apply the chain rule in the reverse order, leading to the following execution trace:

\codespace{}
\begin{center}
\begin{tabular}{r c l c l c l}
$\reversevar{y}$ & = & $\frac{\partial y}{\partial y} $ & = & 1 \\
$\reversevar{v_1}$ & = & $\reversevar{y} \times \frac{\partial y}{\partial v_1}$ & = & $1 \times 1$ & = & 1 \\
$\reversevar{v_2}$ & = & $\reversevar{y} \times \frac{\partial y}{\partial v_2}$ & = & $1 \times 1$ & = & 1 \\
$\reversevar{x_2}$ & = & $\reversevar{v_2} \times \frac{\partial v_2}{\partial x_2}$ & = & $1 \times cos\big(3\big)$ & = & -0.9899 \\
$\reversevar{x_1}$ & = & $\reversevar{v_1} \times \frac{\partial v_1}{\partial x_1}$ & = & $1 \times \frac{1}{1}$ & = & 1 \\
\end{tabular}
\end{center}

\codespace{}

\demo

Forward and reverse mode compute a column and a row, respectively, of the full Jacobian matrix $\mathbf{J}$ at each invocation.
\footnote{$\mathbf{J}\!\mid_{a}$ is a matrix consisting of partial derivatives of the output elements of function $f$ with respect to the elements of the input vector at point $a$.}
More precisely, for a function with an input vector of size $m$ and an output vector of size $n$, the forward mode approach computes a column vector of size $n$, and the reverse mode computes a row vector of size $m$
(see Figure~\ref{jacobian_mat}).


\begin{figure}[t!]
        \centering
        \definecolor{orange}{rgb}{1,0.5,0}
  \begin{equation*}
  f : \mathbb{R}^n \rightarrow \mathbb{R}^m\,\,\, \hspace{0.5cm} \mathbf{J} = 
  \frac{\partial f}{\partial x} = \begin{tikzpicture}[baseline={(m.center)}]
            \matrix [matrix of math nodes,left delimiter={[},right delimiter={]}] (m)
            {
            	\frac{\partial f_1}{\partial x_1} & \cdots & \frac{\partial f_1}{\partial x_m} \\
                \vdots & \ddots & \vdots \\
                \frac{\partial f_n}{\partial x_1} & \cdots & \frac{\partial f_n}{\partial x_m} \\
            };
            \draw[color=blue] (m-1-1.north west) rectangle (m-3-1.south east-|m-1-1.north east) node[pos=1.1,xshift=-0.6cm] {Forward Mode};
            \draw[color=red] (m-1-1.north west) rectangle (m-1-3.south east-|m-1-3.north east) node[pos=1.4,xshift=0.5cm,yshift=0.7cm] {Reverse Mode};
            \draw[color=violet] (-1,1) rectangle (-0.5,-1);
            \draw[color=violet] (-0.3,1) rectangle (-0.2,-1);
            \draw[color=violet] (-0.1,1) rectangle (0,-1);
            \draw[color=violet] (0.1,1) rectangle (0.2,-1);
            \draw[color=violet] (0.5,1) rectangle (1,-1) node[pos=1.1] {\system};
            \end{tikzpicture}
\end{equation*}
        \caption{The Jacobian Matrix of a Function. Forward-mode AD computes a column of this matrix, whereas the reverse-mode AD computes a row of this matrix. \system computes the full Jacobian matrix using a vectorized variant of the forward-mode AD.}
        \label{jacobian_mat}
\end{figure}

From a different point of view, for a given function $f$ with an input vector parameter $a$, forward-mode AD produces the function $df$, where

$$df\;a\;b = \mathbf{J}\!\mid_{a} .\, b$$

\noindent In the case of passing a one-hot vector as $b$, where only the $i^{th}$ element is one, the forward-mode AD computes the $i^{th}$ column of the full Jacobian matrix.
Similarly, for the same function, the reverse-mode AD produces the function $\mathrel{bf}$, where 

$$\mathrel{bf} a\, c = (\mathbf{J}\!\mid_{a})^T .\, c$$

\noindent This expression computes the $j^{th}$ row of the full Jacobian matrix, if $c$ is a one-hot vector with a single one at the $j^{th}$ position and zeros elsewhere.

For a class of
optimization problems, such as various computer vision problems using the Levenberg-Marquardt algorithm~\cite{marquardt1963algorithm,levenberg1944method,more1978levenberg},
one is required to compute the \emph{full} Jacobian matrix.
In such cases, neither of the two techniques perform efficiently,

To compute the full Jacobian matrix, both forward and reverse-mode techniques must iterate either over the columns or the rows of the Jacobian matrix, respectively.
Given that both approaches have a constant overhead over the original computation, the forward mode technique is more efficient for computing the full Jacobian matrix when $n \gg m$, whereas the reverse mode AD is more efficient when $m \gg n$, an uneasy choice.  Moreover:
\begin{itemize}
\item
By carefully examining the body of the loops needed for computing the full Jacobian matrix, one can observe that many computations are loop-invariant and are unnecessarily performed multiple times.
Thus, there is a lost opportunity for \textit{loop-invariant code motion} for hoisting such expressions outside the loop, thus improving the performance (cf. the Bundle Adjustment experiment in Section~\ref{sec:exp}).
\item
Furthermore, while the result of automatic differentiation is known to only have by a constant factor more arithmetic operations than the original program, the constant can be significant; this overhead can have a dramatic impact on the run-time performance in practice. 
More specifically, in applications involving the manipulation of vectors, many intermediate vectors are allocated that can be removed. 
The optimization for eliminating such intermediate vectors is known as \textit{deforestation}~\cite{deforestation,foldr-fusion-1,Svenningsson:2002:SFA:581478.581491,Coutts07streamfusion} or loop fusion in the functional programming community.
This optimization opens the door for many other optimizations such as turning loops iterating over sparse vectors with a single non-zero element into a single statement (cf. Example 4 in Section~\ref{sec:fsmooth_trans}).
\end{itemize}

\subsection{Our contributions}

In this paper, we present a novel automatic differentiation technique based on forward mode, which combines the benefits of both forward and reverse mode in many cases, and which, even for cases that require computing the full Jacobian matrix, outperforms both techniques.  
The key idea behind our technique is that we use a vector-aware programming language, in which the loops required for constructing the full Jacobian matrix are exposed to the compiler.
Thus, the compiler can employ global optimization techniques such as loop-invariant code motion and loop fusion for simplifying the differentiated programs.


\noindent \textbf{Example 1.} Assume that we have a matrix $M$ and two vectors $u$ and $v$ (which are represented as row matrices and are independent of $M$).
Based on matrix calculus one can prove that $\frac{\partial \big(uMv^T\big)}{\partial M}=u^Tv$. 
However, computing the differentiated version of this function using forward-mode AD tools requires multiple iterations over the differentiated program for every element in the matrix $M$. 
By using the reverse-mode AD, one can invoke the differentiated function only once, and the adjoin parts of the input matrix $M$ will be filled in.
We show in Section~\ref{sec:fsmooth_trans} that \system{} derives the gradient of this expression with respect to $M$, resulting in an expression equivalent to $u^Tv$.
This removes the need for multiple iterations over the differentiated program for each element of matrix $M$, in contrast to the existing AD tools based on the forward-mode AD technique.



The contributions of this paper are summarized as follows:
\begin{itemize}
\item We present \fsmooth, a higher-order functional array-processing language in Section~\ref{sec:fsmooth}. This language can be efficiently compiled into low-level C code with efficient memory management. Then, we present \ladsl{}, a linear algebra DSL inspired by MATLAB, embedded~\cite{hudak-dsl} in this language in Section~\ref{sec:ladsl}.
\item Then, we show the differentiation programming capabilities provided by \fsmooth. 
First, Section~\ref{sec:diff_api} shows the high-level API exposed in \fsmooth{} for performing various matrix derivatives such as scalar derivatives, gradients, and Jacobians.
Then, we show transformation rules for performing source-to-source automatic differentiation of \fsmooth{} expressions in Section~\ref{sec:fsmooth_ad}.
\item
Afterwards, we show how \system produces efficient differentiated programs by introducing several global optimizations
such as loop-invariant code motion, loop fusion, and partial evaluation, as well as generating C code with efficient stack-discipline memory management in Section~\ref{sec:fsmooth_trans}.
\item Finally, using several micro benchmarks and several functions used in machine learning and computer vision workloads, we show how \system outperforms the state-of-the-art AD techniques in Section~\ref{sec:exp}.
\end{itemize}




%% file: overview.tex
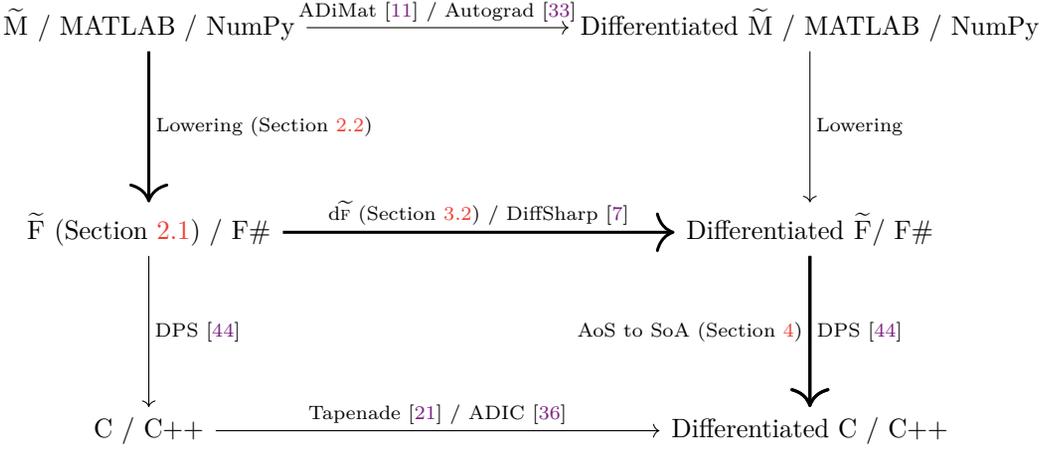
\begin{figure}[t!]
\begin{equation*}
\begin{tikzcd}[column sep=3.5cm, row sep=2cm]
\text{\ladsl{} / MATLAB / NumPy} \arrow[d,line width=0.4mm,"\text{Lowering (Section \ref{sec:ladsl})}"] \arrow[r,"\text{ADiMat~\cite{bischof2002combining} / Autograd~\cite{maclaurin2015autograd}}"] & \text{Differentiated \ladsl{} / MATLAB / NumPy} \arrow[d,"\text{Lowering}"] \\
\text{\fsmooth~(Section~\ref{sec:fsmooth}) / F\#} \arrow[r,line width=0.4mm,"\text{\system{} (Section~\ref{sec:fsmooth_ad}) / DiffSharp~\cite{baydin2015diffsharp}}"]  \arrow[d,"\text{DPS~\cite{dps_fhpc}}"]  & \text{Differentiated \fsmooth / F\#} \arrow[d,line width=0.4mm,swap,"\text{AoS to SoA (Section \ref{sec:fsmooth_trans})}"] \arrow[d,line width=0.4mm,"\text{DPS~\cite{dps_fhpc}}"] \\
\text{C / C++}  \arrow[r,"\text{Tapenade~\cite{tapenade} / ADIC~\cite{narayanan2010adic2}}"]    & \text{Differentiated C / C++} \\
\end{tikzcd}
\end{equation*}
\caption{Compilation process in \system and other AD systems. The solid arrows correspond to the pipeline used in \system.}
\label{fig:comp_cd}
\end{figure}

\section{Overview}
\label{sec:overview}

In this section, we start with an overview of the compilation process in \system, which is shown in Figure~\ref{fig:comp_cd}. This figure demonstrates the position of \system with respect to existing AD tools.
\system starts from a program written in a high-level linear algebra DSL, called \ladsl{} (Section~\ref{sec:ladsl}).
This program is lowered into its implementation in a higher-order functional language with array support, called \fsmooth{} (Section~\ref{sec:fsmooth}).
If a part of the program requires computing differentiation (which are specified by using high-level differentiation API exposed by \system, as mentioned in Section~\ref{sec:diff_api})
\system uses AD transformation rules (Section~\ref{sec:fsmooth_ad}) for transforming the involved expressions into their differentiated form.
Finally, after applying several simplifications such as loop fusion, partial evaluation, data layout transformation, etc. (Section~\ref{sec:fsmooth_trans}) the differentiated program is transformed into low-level C code.
The generated C code uses efficient stack-discipline memory management by using the destination-passing style (DPS) technique~\cite{dps_fhpc}. 

Next, we present the core functional language used in \system, on top of which we define source-to-source AD transformation and simplification rules.

\subsection{\fsmooth}
\label{sec:fsmooth}

\lafsharp{} 
(we pronounce it F smooth) 
is a subset of F\#, an ML-like functional programming language.  It is designed to be \emph{expressive enough} to make it easy to write array-processing workloads, while simultaneously being \emph{restricted enough} to be able to define automatic differentiation rules and allow it to be compiled to code that is as efficient as hand-written C, with very simple and efficient memory management~\cite{dps_fhpc}. 

\begin{figure}[t]
\input{form/fsmooth-syntax}
\caption{The syntax, type system, and function constants of the core \fsmooth{}.}
\label{fig:fsmooth_core_syntax}
\end{figure}

Figure~\ref{fig:fsmooth_core_syntax} shows the abstract syntax (parentheses can be used as necessary), type system, and several built-in functions of \fsmooth. 
In addition to the usual $\lambda$-calculus constructs (abstraction, application, and variable access), \fsmooth{} supports let binding and conditionals.

\fsmooth{} supports array programming by defining the following built-in functions: \vbuildk{} for producing arrays;  \viteratek{} for iteration for a particular number of times (from \cod{0} to \cod{n-1}) while maintaining a state across iterations; \vclength{} to get the size of an array; and \vcget{} to index an array. 

One of the key features of \fsmooth{} is its support for both source-to-source automatic differentiation and global optimizations such as loop-invariant code motion and loop fusion in the same time.
The transformations required for automatic differentiation are presented in Section~\ref{sec:fsmooth_ad}, and the ones for optimization and simplification are shown in Section~\ref{sec:fsmooth_trans}.

Next, we show how a Linear Algebra DSL can be defined on top of \fsmooth{}.

\subsection{\ladsl{}}
\label{sec:ladsl}
\ladsl{} is a functional Linear Algebra DSL, mainly inspired by MATLAB and R, programming languages which are heavily used by data analysts.
By providing high-level vector and matrix operations, \ladsl{} frees the users from low-level details and enables them to focus on the algorithmic aspects of the problem in hand. 

\begin{table}[t!]
\centering
\begin{tabular}{ c | c | c | c }
  Matlab & R & NumPy & \ladsl{}  \\
  \hline			
  A * B & A \%*\% B & A.dot(B) & matrixMult A B \\
  A + B & A + B & A + B & matrixAdd A B \\
  A' & t(A) & A.T & matrixTranspose A \\
  ones(n, m) & matrix(1, n, m) & ones((n, m)) & matrixOnes n m \\
  zeros(n, m) & matrix(0, n, m) & zeros((n, m)) & matrixZeros n m \\
  eye(n) & diag(n) & eye(n) & matrixEye n \\
  
\end{tabular}
\vspace{0.3cm}
\caption{Equivalent operations in Matlab, R, NumPy, and \ladsl{}.}
\label{table_ops}
\end{table}

\ladsl{} is an \textit{embedded DSL} (EDSL)~\cite{hudak-dsl} in \fsmooth{}; it is defined as a library on top of \fsmooth{}.
Figure~\ref{fig:ladsl_ops} demonstrates a subset of \ladsl{} operations which are defined as functions in \fsmooth{}. 
This DSL is expressive enough for constructing vectors and matrices, elementwise-operations, accessing a slice of elements, reduction-based operations (computing the sum of vector elements), matrix transpose, and matrix multiplication.
Supporting more sophisticated operations such as matrix determinant and matrix decomposition is beyond the scope of the current paper, and we leave it for the future.
As discussed before, \ladsl{} is inspired by MATLAB and R. 
As a result, there is a mapping among the constructs of \ladsl{} and these matrix-based languages. 
Hence, it is easily possible to translate a program written in one of these languages to \ladsl{}.
Table~\ref{table_ops} demonstrates the mapping among a subset of the constructs of MATLAB, R, NumPy and \ladsl{}.

\input{form/fsmooth-lib}

\noindent \textbf{Example 1 (Continued).} The matrix expression $uMv^T$ is expressed as the following function in \ladsl{}:

\codespace{}
\begin{tabular}{l}
\lett{} f = \vabs{u M v}{}\\
\tabt{}\lett{} um = vectorToMatrix u\\
\tabt{}\lett{} vt = matrixTranspose (vectorToMatrix v)\\
\tabt{}\lett{} m = 
matrixMult um (matrixMult M
vt)\\
\tabt{}\vget{\vget{m}{0}}{0}
\end{tabular}

\codespace{}
\noindent The last expression is for accessing the single scalar element of a $1\times 1$ matrix.

\demo

%% file: form/fsmooth-syntax.tex
\raggedright
\def\comment{ & -- }
\begin{tabular*}{\columnwidth}{r c l r}
\expr & ::= & \expr{} \expr{} | \vabs{\text{x}}{\expr} | \text{x} \comment Application, Abstraction, and Variable Access\\
& | & \text{n} | \text{i} |  \valcard{} \comment Scalar, Index, and Cardinality Value\\
& | & \text{c} \comment Constants (see below)\\
& | & \lett{} x = \expr{} \inn{} \expr{} \comment(Non-Recursive) Let Binding\\
& | & \vifthenelse{\expr{}}{\expr{}}{\expr{}} \comment Conditional\\
\typet{} & ::= & \typemat{} \comment (Non-Functional) Expression Type\\
& | & \typefunone{\typet}{\typet} \comment Function Types\\
\typemat{} & ::= & \typenum{} \comment Numeric Type\\
& | & \typearray{\typemat{}} \comment Vector, Matrix, ... Type\\
& | & \typepair{\typemat}{\typemat} \comment Pair Type \\
& | & \typebool \comment Boolean Type\\
\typenum & ::= & \typedouble{} | \typeindex{} | \typecard{} \comment Scalar, Index, and Cardinality Type\\
\end{tabular*}\\[8pt]
\textbf{Typing Rules:}\\
\begin{minipage}{\columnwidth}
\centering
(T-App) $\infer{\exprind{1}: \typefunone{\typeind{1}}{\typeind{2}} \tab \exprind{2} : \typeind{1}}{\exprind{1}\ \exprind{2}: \typeind{2}}$
(T-Abs) $\infer{ \Gamma  \cup  \text{x}: \typeind{1} \vdash \expr{}: \typeind{2} }{\Gamma \vdash \vabs{\text{x}}{\expr}: \typefunone{\typeind{1}}{\typeind{2}}}$
(T-Var) $\infer{\text{x}: \typet \in \Gamma }{\Gamma \vdash \text{x}: \typet}$
\\
(T-Let) $\infer{ \Gamma \vdash \exprind{1}: \typeind{1} \tab \tab \Gamma, \text{x}: \typeind{1} \vdash \exprind{2}: \typeind{2}}{\Gamma \vdash \text{\lett{} x = \exprind{1} \inn{} \exprind{2}: \typeind{2}}}$ 
\tab (T-If) $\infer{\exprind{1}: \typebool{} \tab \tab \exprind{2}: \typemat \tab \tab \exprind{3}: \typemat }{\vifthenelse{\exprind{1}}{\exprind{2}}{\exprind{3}}: \typemat}$ 
\end{minipage}\\
\textbf{Scalar Function Constants:}\\
\begin{minipage}{0.48\columnwidth}
\begin{tabular}{l c l}
\vconst{+} | \vconst{-} | \vconst{*} | \vconst{/} | \vconst{**} &:& \typefunone{\typenum, \typenum}{\typenum}\\
\vconst{sin} | \vconst{cos} | \vconst{tan} |\\ \vconst{log} | \vconst{exp} &:& \typefunone{\typenum}{\typenum}\\
\end{tabular}
\end{minipage}
\begin{minipage}{0.40\columnwidth}
\begin{tabular}{l c l}
\vconst{>} | \vconst{<} | \vconst{==} | \vconst{<>} &:& \typefunone{\typenum{} \funarrow{} \typenum}{\typebool} \\
\vconst{\&\&} | \vconst{||} &:& \typefunone{\typebool{} \funarrow{} \typebool}{\typebool} \\
\vconst{!} &:& \typefunone{\typebool}{\typebool} \\
\end{tabular}
\end{minipage}\\[8pt]
\textbf{Vector Function Constants:}\\
\begin{minipage}{0.57\columnwidth}
\setlength\tabcolsep{1.5pt} 
\begin{tabular}{l c l}
\vbuildk{} &:& 
\typefunone{
\typecard{}
 \funarrow{}
 (\typefunone{\typeindex}{\typemat{}})
 }{
 \typearray{\typemat}
}\\
\viteratek{} &:& 
\typefunone{%
(\typefunone{%
$\typemat$
\funarrow{}
\typeindex
}{$\typemat$})
\funarrow{}
$\typemat$
\funarrow{}
\typecard
}{$\typemat$
}\\
\end{tabular}
\end{minipage}
\begin{minipage}{0.40\columnwidth}
\setlength\tabcolsep{1.5pt} 
\begin{tabular}{l c l}
\vcget{} &:& 
\typefunone{%
\typearray{\typemat}%
\funarrow{}%
\typeindex%
}{%
\typemat%
}\\
\vclength{} &:& 
\typearray{\typemat}%
\funarrow{}%
\typecard\\
\end{tabular}
\end{minipage}\\[8pt]
\textbf{Pair Function Constants:}\\
\vpairk{} : \typefunone{%
$\typemat{}_1$%
}{%
\typefunone{%
$\typemat{}_2$%
}{\typepair{$\typemat{}_1$}{$\typemat{}_2$}}
}\tab\tab\tab
\vfstk{} : \typefunone{%
\typepair{$\typemat{}_1$}{$\typemat{}_2$}%
}{%
$\typemat{}_1$%
}\tab\tab\tab
\vsndk{} : \typefunone{%
\typepair{$\typemat{}_1$}{$\typemat{}_2$}%
}{%
$\typemat{}_2$%
}
\\[8pt]
\textbf{Syntactic Sugar:} \\
\begin{minipage}{0.38\columnwidth}
\begin{tabular}{l c l}
\vget{\exprind{0}}{\exprind{1}} & = & \vcget{} \exprind{0} \exprind{1}
\\
\vpair{\exprind{0}}{\exprind{1}} & = & \vpairk{} \exprind{0} \exprind{1} \\
\exprind{1} $bop$ \exprind{2} & = & $bop$ \exprind{1} \exprind{2} \\
\typevector{} & = & \typearray{\typedouble}
\end{tabular}
\end{minipage}
\begin{minipage}{0.60\columnwidth}
\begin{tabular}{l c l}
\typematrix{} & = & \typearray{\typearray{\typedouble}} \\
\typedoubled{} & = & \typepair{\typedouble}{\typedouble} \\
\typevectord{} & = & \typearray{\typepair{\typedouble}{\typedouble}} \\
\typematrixd{} & = & \typearray{\typearray{\typepair{\typedouble}{\typedouble}}}
\end{tabular}
\end{minipage}

%% file: form/fsmooth-lib.tex
\begin{figure*}[t]
\hfill\begin{minipage}{.49\textwidth}\raggedright
\lett{} vectorRange = \vabs{n}{}
\\ \tabt 
\vbuild{n}{(\vabs{i}{i})}
\\
\lett{} vectorFill = \vabs{n e}{} 
\\ \tabt
\vbuild{n}{(\vabs{i}{e})}
\\
\lett{} vectorHot = \vabs{n i}{} 
\\ \tabt
\vbuild{n}{(\vabs{j}{\iif{} i = j \then{} 1 \elsee{} 0})}
\\
\lett{} vectorMap = \vabs{v f}{}
\\ \tabt
\vbuild{(\vlength{v})}{(\vabs{i}{f \vget{v}{i}})}
\\
\lett{} vectorMap2 = \vabs{v1 v2 f}{}
\\
\tabt 
\vbuild{(\vlength{v1})}{(\vabs{i}{f \vget{v1}{i} \vget{v2}{i}})}
\\
\lett{} vectorZip = \vabs{v1 v2}{}
\\
\tabt 
vectorMap2 v1 v2 (\vpairk{})
\\
\lett{} vectorAdd = \vabs{v1 v2}{}
\\
\tabt 
vectorMap2 v1 v2 (+)
\\
\lett{} vectorEMul = \vabs{v1 v2}{}
\\
\tabt 
vectorMap2 v1 v2 ($\times$)
\\
\lett{} vectorSMul = \vabs{v s}{}
\\
\tabt 
vectorMap v (\vabs{a}{a $\times$ s})
\\
\lett{} vectorSum = \vabs{v}{}\\
\tabt \viterate{(\vabs{s i}{s + \vget{v}{i}})}{0}{(\vlength{v})}
\\
\lett{} vectorDot = \vabs{v1 v2}{}\\
\tabt
vectorSum (vectorEMul v1 v2)
\\
\lett{} vectorNorm = \vabs{v}{}
\\
\tabt 
sqrt (vectorDot v v)
\\
\lett{} vectorSlice = \vabs{v s e}{} \\
\tabt \vbuild{(e \vcsubcard{} s \vcaddcard{} 1)}{(\vabs{i}{\vget{v}{i + s}})}
\\
\lett{} vectorToMatrix = \vabs{v}{}
\\
\tabt \vbuild{1}{(\vabs{i}{v})} 
\\
\lett{} vectorOutProd = \vabs{v1 v2}{}
\\
\tabt \lett{} m1 = vectorToMatrix v1 
\\
\tabt \lett{} m2 = vectorToMatrix v2 
\\
\tabt \lett{} m2T = matrixTranspose m2
\\
\tabt matrixMul m1 m2T
\end{minipage}\hfill\begin{minipage}{.49\textwidth}\raggedright
\lett{} matrixRows = \vabs{m}{} \vlength{m} 
\\
\lett{} matrixCols = \vabs{m}{} \vlength{(\vget{m}{0})} 
\\
\lett{} matrixZeros = \vabs{r c}{} 
\\ \tabt \vbuild{r}{(\vabs{i}{vectorFill c 0})}
\\
\lett{} matrixOnes = \vabs{r c}{} 
\\ \tabt \vbuild{r}{(\vabs{i}{vectorFill c 1})}
\\
\lett{} matrixEye = \vabs{n}{} 
\\ \tabt \vbuild{n}{(\vabs{i}{vectorHot n i})}
\\
\lett{} matrixHot = \vabs{n m r c}{} 
\\ \tabt
\vbuild{n}{(\vabs{i}{}} 
\\ \tabt \tabt
\vbuild{m}{(\vabs{j}{}} 
\\ \tabt \tabt \tabt \iif{} (i = r \cod{\&\&} j = c) \then{} 1 \elsee{} 0
\\
\tabt ) )
\\
\lett{} matrixMap = \vabs{m f}{}
\\ \tabt
\vbuild{(\vlength{m})}{(\vabs{i}{f \vget{m}{i}})}
\\
\lett{} matrixMap2 = \vabs{m1 m2 f}{}
\\
\tabt 
\vbuild{(\vlength{m1})}{(\vabs{i}{f \vget{m1}{i} \vget{m2}{i}})}
\\
\lett{} matrixAdd = \vabs{m1 m2}{}
\\
\tabt
matrixMap2 m1 m2 vectorAdd
\\
\lett{} matrixTranspose = \vabs{m}{} 
\\
\tabt
\vbuild{(matrixCols m)}{}(\vabs{i}{}
\\
\tabt \tabt 
\vbuild{(matrixRows m)}{}(\vabs{j}{}
\\
\tabt \tabt \tabt 
\vget{\vget{m}{j}}{i}
\\
\tabt ) )
\\
\lett{} matrixMul = \vabs{m1 m2}{} 
\\
\tabt \lett{} m2T = matrixTranspose m2
\\
\tabt
\vbuild{(matrixRows m1)}{}(\vabs{i}{}
\\
\tabt \tabt \vbuild{(matrixCols m2)}{}(\vabs{j}{}
\\
\tabt \tabt \tabt vectorDot (\vget{m1}{i}) (\vget{m2T}{j})
\\
\tabt ) )
\\
\lett{} matrixTrace = \vabs{m}{}\\
\tabt{}\vifoldk{}\,(\vabs{s i}{}%
s+\vget{\vget{m}{i}}{i}) 0 (\vlengthk{}\,m)\\
\end{minipage}\hfill
\caption{A subset of \ladsl{} constructs defined in \lafsharp{}.}
\label{fig:ladsl_ops}
\end{figure*}

%% file: fsmooth_d.tex
\section{Differentiation}
\label{sec_diff}

In this section, we show the differentiation process in \system. First, we start by the high-level API exposed by \system to the end users.
Then, we show how \system uses automatic differentiation behind the scenes for computing derivatives.
Finally, we present the optimizations offered by \system, and we demonstrate how \system can use these optimizations to deduce several matrix calculus identities.

\subsection{High-Level API}
\label{sec:diff_api}
For computing the derivative of an arbitrary function, \system provides the \derivk{} construct. 
This construct can be better thought of as a macro, 
which is expanded during compilation time.
The expanded expression includes the expression of the original computation, which is given as the first argument (and can be an arbitrary scalar, vector, or matrix expression), and the derivative of this expression with respect to the variable given as the second argument, referred to as the \textit{independent variable}.
Note that one can easily compute the derivative of an expression with respect to a list of free variables by multiple invocation of the \derivk{} construct.

Algorithm~\ref{fig:deriv} shows a pseudo-code implementation of the \derivk{} construct.
First, \derivk{} constructs a lambda function which has the free variables of the given expression as its input parameters (cf. line 6). 
This function is given as input to source-to-source automatic differentiation for computing the derivative (cf. line 8).
The differentiated function is applied to the dual number encoding of all the free variables (cf. lines 5-8). 
If the free variable is different than the input variable with respect to which we are differentiating (i.e., the independent variable), the derivative part is a zero scalar, vector, or matrix (cf. lines 26-33). 
Otherwise, the derivative part is a one-hot encoding scalar, vector, or matrix (cf. lines 35-42).

If the independent variable has a scalar type, \derivk{} returns the applied function (cf. lines 9-11). 
However, if the independent variable has a vector type, \derivk{} constructs a vector with the same number of elements as the independent variable. 
For computing the $ri^{th}$ element of the result vector, the corresponding input vector is a one-hot encoding with a single one at the $ri^{th}$ position (cf. lines 12 and 39).
The situation is similar for an independent variable with a matrix type; the corresponding one-hot encoding matrix has a single one at the $ri^{th}$ row and $ci^{th}$ column (cf. lines 14 and 41). Note that the two variables ri and ci are treated specially and are distinguished variables. 

\noindent \textbf{Example 2.} Let us assume that we would like to compute the derivative of a program computing the cosine function with respect to its input:

\codespace{}
cos(a)

\codespace{}
\noindent The derivative of this program at point $a$ is represented as follows:

\codespace{}
\dterm{\deriv{(cos a)}{a}}

\codespace{}
\noindent This expression is transformed into the following expression after expanding the \derivk{} macro:

\codespace{}
\dterm{(\difftrans{\vabs{a}{}cos(a)}) \adpair{a}{1}}

\demo

\begin{algorithm}[t!]
\renewcommand{\algorithmiccomment}[1]{// #1}
\begin{algorithmic}[1]
\State \Comment{Returns an expression including both the original and the derivative computation.}
\Function{deriv}{\expr{}, \text{x}}
\State args $\gets \emptyset$
\State f $\gets$ \expr{}
\ForAll{v $\gets$ \textsc{freeVars}(e)}
\State f $\gets$ \vabs{v}{f}
\State  args $\gets$ args $\cup$ \Call{dual}{v, \iif{}(v = x) \then{} \textsc{oneHot}(v) \elsee{} \textsc{zero}(v)}
\EndFor
\State df $\gets$ (\difftrans{f}) args
\If{\Call{Type}{x} = \typedouble}
\State \Return{df}
\ElsIf{\Call{Type}{x} = \typevector}
\State \Return{\vbuildk{} (\vlength{x}) (\vabs{ri}{} df)}
\ElsIf{\Call{Type}{x} = \typematrix}
\State \Return{\vbuildk{} (matrixRows x) (\vabs{ri}{} \vbuildk{} (matrixCols x) (\vabs{ci}{} df))}
\EndIf
\EndFunction
\State \Comment{Returns the dual number encoding of the two input expressions.}
\Function{dual}{\exprind{1}, \exprind{2}}
\If{\Call{Type}{\exprind{1}} = \typedouble}
\State \Return{\adpair{\exprind{1}}{\exprind{2}}}
\ElsIf{\Call{Type}{\exprind{1}} = \typevector}
\State \Return{vectorZip \exprind{1} \exprind{2}}
\ElsIf{\Call{Type}{\exprind{1}} = \typematrix}
\State \Return{matrixZip \exprind{1} \exprind{2}}
\EndIf
\EndFunction
\State \Comment{Returns a zero scalar, vector, or matrix expression based on the type of input.}
\Function{zero}{\expr}
\If{\Call{Type}{\expr} = \typedouble}
\State \Return{0}
\ElsIf{\Call{Type}{\expr} = \typevector}
\State \Return{vectorZeros (\vlength{\expr})}
\ElsIf{\Call{Type}{\expr} = \typematrix}
\State \Return{matrixZeros (matrixRows \expr) (matrixCols \expr)}
\EndIf
\EndFunction
\State \Comment{Returns a one-hot encoding scalar, vector, or matrix expression.}
\Function{oneHot}{\expr}
\If{\Call{Type}{\expr} = \typedouble}
\State \Return{1}
\ElsIf{\Call{Type}{\expr} = \typevector}
\State \Return{vectorHot (\vlength{\expr}) ri}
\ElsIf{\Call{Type}{\expr} = \typematrix}
\State \Return{matrixHot (matrixRows \expr) (matrixCols \expr) ri ci}
\EndIf
\EndFunction
\end{algorithmic}
\caption{A pseudo-code implementation of the \derivk{} construct.}
\label{fig:deriv}
\end{algorithm}

Furthermore, \system provides three additional differentiation constructs, inspired by AD tools such as DiffSharp~\cite{baydin2015diffsharp}: 1) \diffk{} computes the derivative a function, from a real number to a real number, with respect to its input, 
2) \gradk{} computes the gradient of a function, from a vector of real numbers to a real number, with respect to its input vector,
and 3) \jacobk{} computes the Jacobian matrix of a vector-valued function, a function from a vector of real numbers to a vector of real numbers, with respect to its input vector.
Figure~\ref{fig:diff_trans_api} demonstrates how these high-level differentiation constructs are defined in terms of the source-to-source AD transformation construct $\mathcal{D}$.

\input{form/fsmooth-diff-api}

\noindent \textbf{Example 2 (Continued).} 
For the previous example, if we would like to use the \diff{} construct, first we have to define the following function:

\codespace{}
g = \vabs{x}{}cos(x)

\codespace{}
\noindent The derivative of this function at point $a$ is represented as follows:

\codespace{}
\dterm{(\diff{g})~a}

\codespace{}
\noindent which is expanded to the following program:

\codespace{}
\dterm{\difftrans{g} \adpair{a}{1}}

\demo

\noindent Table~\ref{tbl:matrix_derivative} summarizes different matrix derivatives, and how they can be computed using our high-level API. Note that the definition of \vdiffk{} and \mdiffk{} is similar to \diffk, and the definition of \mgradk{} is similar to \gradk{} and \jacobk{} (cf. Figure~\ref{fig:diff_trans_api}). 
Note that the \derivk{} construct subsumes all these operators.   

\begin{table}
\begin{tabular}{c | c | c | c}
\backslashbox{Input Type}{Output Type} & Scalar & Vector & Matrix \\ \hline
Scalar & \diffk{} & \vdiffk{} & \mdiffk{}  \\ \hline
Vector & \gradk{} & \jacobk{} & -- \\ \hline
Matrix & \mgradk{} & -- & --
\end{tabular}
\caption{Different types of matrix derivatives.}
\label{tbl:matrix_derivative}
\end{table}

One key advantage of defining different matrix derivatives in terms of automatic differentiation is that one no longer needs to define the matrix calculus derivative rules for all different combinations shown in Table~\ref{tbl:matrix_derivative}.
Instead these rules can be deduced automatically from the automatic differentiation rules defined for scalar values. 
Moreover, even the algebraic identities for matrix derivative can be deduced by using the simplification rules presented in Section~\ref{sec:fsmooth_trans}.

Next, we present the source code transformation required for applying automatic differentiation rules.
\subsection{Source-to-Source Automatic Differentiation}
\label{sec:fsmooth_ad}

\system relies on source-to-source translation for implementing forward-mode automatic differentiation.
Each expression is converted into an expression containing both the original computation, together with the derivative computation, a.k.a. the dual number technique.
The scalar expressions are transformed into a pair of values, the original computation and the derivative computation.
The vector expressions are transformed into vectors containing tuple expressions, instead of scalar expressions.
The situation is similar for higher-rank tensors such as matrices.

The rules for automatic differentiation are demonstrated in Figure~\ref{fig:diff_trans}. 
\difftrans{e} specifies the AD translation for expression e. 
A variable y is translated as \diffvarprefix{\text{y}}, emphasizing that the translated variable keeps the derivative part as well (D-Abs, D-Var, and D-Let).
\difftransp{\expr} is a shorthand for extracting the original computation from the translated term \difftrans{\expr}, while \difftranst{\expr} is a shorthand for accessing the derivative part.
In the case of scalar expressions, \difftransp{\expr} and \difftranst{\expr} are equivalent to accessing the first and the second element of the result dual number, respectively (D-NumV and D-NumT).

Constructing an array is differentiated as an array with the same size, however, the way that each element of the array is constructed is differentiated (D-Build).
Differentiating an iteration results in an iteration with the same number of iterations, and with the initial state and the next state function both differentiated (D-IFold).
The differentiation of the length and indexing an array, is the same as the length and indexing the differentiated array, respectively (D-Length and D-Get).

Differentiating a pair of elements results in the pair of differentiated elements (D-Pair). Similarly, differentiating the projection of a pair, is the projection of the differentiated pair (D-Fst, D-Snd).
For other scalar-valued functions, the differentiation rules are
similar to the corresponding rules in mathematics.

\noindent \textbf{Example 2 (Continued).} In the previous example, based on the automatic differentiation rules, the differentiated program would be as follows:

\codespace{}
\diffvarprefix{g} = \vabs{\diffvarprefix{x}}{}-\dterm{\diffvarprefix{x}} * sin(\pterm{\diffvarprefix{x}})

\codespace{}
\noindent Based on the definition of the \diff{} construct, we have to use the AD version of the function (i.e., g) and assign 1 to the derivative part of the input. So the value of $cos'$ for the input a is computed as follows:

\codespace{}
\dterm{(\diff{g})~a} \tab \evalsto \tab 
\dterm{\difftrans{g} \adpair{a}{1}} \tab \evalsto \tab  
\dterm{\diffvarprefix{g} \adpair{a}{1}} \tab \evalsto 

\tab
\text{-\dterm{\adpair{a}{1}} * sin(\pterm{\adpair{a}{1}})} \tab \evalsto \tab 
\text{-1 * sin(a)} \tab \evalsto \tab  
\text{-sin(a)}

\demo

\noindent Similarly, we can compute the partial derivatives of a given function, by setting the desired derivative part to one, and the rest of derivatives to zero. This process is illustrated in the next example.

\noindent \textbf{Example 3.} Assume that we would like to compute the partial derivative of the expression a \vconst{*} b with respect to a, which is represented as follows in \fsmooth{}:

\codespace{}
\dterm{\deriv{(a \vconst{*} b)}{a}}

\codespace{}
\noindent This expression is expanded as follows:

\codespace{}
\dterm{\difftrans{\vabs{a b}{a \vconst{*} b}} \adpair{a}{1} \adpair{b}{0}}

\codespace{}
\noindent Note that the derivative part of the second input is set to 0. Similar to the previous example, the result is as follows:

\codespace{}
\dterm{(\vabs{\diffvarprefix{a}~\diffvarprefix{b}}{}
\adpair{\pterm{\diffvarprefix{a}}\vconst{*}\pterm{\diffvarprefix{b}}}{\pterm{\diffvarprefix{a}}\vconst{*}\dterm{\diffvarprefix{b}} +
\dterm{\diffvarprefix{a}}\vconst{*}\pterm{\diffvarprefix{b}}}) \adpair{a}{1} \adpair{b}{0}}

\codespace{}
\noindent which is evaluated as follows:

\codespace{}
\dterm{\adpair{a \vconst{*}  b}{1 \vconst{*}  b + a \vconst{*}  0}} \tab \evalsto \tab
1 \vconst{*} b + a \vconst{*} 0 \tab \evalsto \tab
b

\demo

\input{form/fsmooth-diff}

\noindent It is important to note that \system performs many of the evaluation steps shown for the previous examples during compilation time, i.e., \system performs partial evaluation. 

\input{perturb}

Next, we give more details on the optimizations and simplifications offered by \system.

%% file: form/fsmooth-diff-api.tex
\begin{figure}[t]
\centering
\begin{tabular}{l | l | l}
Oper. & Type & Definition\\ \hline
\diff{} & (\typedouble \funarrow \typedouble) \funarrow &
\vabs{f x}{}\difftrans{f} \adpair{x}{1}\\
& \tab \typedouble \funarrow \typedoubled \\ \hline
\grad{} & (\typevector \funarrow \typedouble)  & 
\vabs{f v}{}
\\ & \tab \funarrow \typevector \funarrow \typevectord &  
\tab \vbuildk{} (\vlengthk~v) (\vabs{i}{}\\ \cline{1-2}
\jacob{} & (\typevector \funarrow \typevector)  &
\tab \tab \difftrans{f} (vectorZip v (vectorHot (\vlengthk~v) i))
\\ & \tab \funarrow \typevector \funarrow \typematrixd &  
\tab )\\
\end{tabular}
\caption{High-Level Differentiation API for \lafsharp{}.}
\label{fig:diff_trans_api}
\end{figure}

%% file: form/fsmooth-diff.tex
\begin{figure*}[t!]
\centering
\setlength{\tabcolsep}{2pt}
\begin{tabular}{l r c l}
(D-App) & 
\difftrans{\vapp{\exprind{0}}{\exprind{1}}}&=&
(\difftrans{\exprind{0}})\ (\difftrans{\exprind{1}})\\
(D-Abs) &
\difftrans{\vabs{\text{x}}{\expr{}}}&=&
\vabs{\diffvarprefix{\text{x}}}{ \difftrans{\expr{}} } \\
(D-Var) & 
\difftrans{\text{y}}&=&
\diffvarprefix{\text{y}}\\
(D-Let) &
\difftrans{\vlet{\text{x}}{\exprind{1}}{\exprind{2}}}&=&
\vlet{\diffvarprefix{\text{x}}}{\difftrans{\exprind{1}}}{}\\
& & & 
\tabt \difftrans{\exprind{2}}
\\
(D-If) & 
\difftrans{\vifthenelse{\exprind{1}}{\exprind{2}}{\exprind{3}}}&=&
\vifthenelse{(\vfstk{} \difftrans{\exprind{1}})}{\difftrans{\exprind{2}}}{\difftrans{\exprind{3}}}\\
\\
(D-Build) & \difftrans{\vbuild{\exprind{0}}{\exprind{1}}} &=&
\vbuild{(\vfstk{} \difftrans{\exprind{0}})}{(\vabs{i}{(\difftrans{\exprind{1}}) (i, 0))}}
\\
(D-IFold) & \difftrans{\viterate{\exprind{0}}{\exprind{1}}{\exprind{2}}} &=&
\viteratek{} (\vabs{x i}{} 
\\
& & & \tabt 
(\difftrans{\exprind{0}}) x (i, 0)%
) \difftrans{\exprind{1}} (\vfstk{} \difftrans{\exprind{2}})\\
(D-Get) & \difftrans{\vget{\exprind{0}}{\exprind{1}}} &=&
\vget{(\difftrans{\exprind{0}})}{\vfstk{} \difftrans{\exprind{1}}}
\\
(D-Length) & \difftrans{\vlength{\exprind{0}}} &=&
(\vlength{\difftrans{\exprind{0}}}, 0)
\\
\\
(D-Pair) & \difftrans{\vpair{\exprind{0}}{\exprind{1}}} &=&
\vpair{\difftrans{\exprind{0}}}{\difftrans{\exprind{1}}}
\\
(D-Fst) & \difftrans{\vfstk{} \exprind{0}} &=&
\vfstk{} (\difftrans{\exprind{0}})
\\
(D-Snd) & \difftrans{\vsndk{} \exprind{0}} &=&
\vsndk{} (\difftrans{\exprind{0}})
\\
\\
(D-NumV) & \expr: \typenum{} $\vdash$ \difftransp{\expr} &=&
\vfstk{} \difftrans{\expr}
\\
(D-NumT) & \expr: \typenum{} $\vdash$ \difftranst{\expr} &=&
\vsndk{} \difftrans{\expr}
\\
\\
(D-Neg) & \difftrans{-\exprind{1}} &=&
\adpair{
-\difftransp{\exprind{1}}
}{
-\difftranst{\exprind{1}}
}
\\
(D-Add) & \difftrans{\exprind{1} + \exprind{2}} &=&
\adpair{
\difftransp{\exprind{1}}+\difftransp{\exprind{2}}
}{
\difftranst{\exprind{1}}+\difftranst{\exprind{2}}
}
\\
(D-Mult) & \difftrans{\exprind{1} * \exprind{2}} &=&
\adpair{
\difftransp{\exprind{1}}*\difftransp{\exprind{2}}
}{
\difftranst{\exprind{1}}*\difftransp{\exprind{2}} + \difftransp{\exprind{1}}*\difftranst{\exprind{2}}
}
\\
(D-Div) & \difftrans{\exprind{1} / \exprind{2}} &=&
(
\difftransp{\exprind{1}}/\difftransp{\exprind{2}} 
,
\\
& & &
\,\,
(\difftranst{\exprind{1}}*\difftransp{\exprind{2}} - \difftransp{\exprind{1}}*\difftranst{\exprind{2}}) / (\difftransp{\exprind{2}}**2)
)
\\
(D-Pow) & \difftrans{\exprind{1} ** \exprind{2}} &=&
(
\difftransp{\exprind{1}}**\difftransp{\exprind{2}} 
,
(\difftransp{\exprind{2}} * \difftranst{\exprind{1}} / \difftransp{\exprind{1}} +
\\
& & &
\,\,\,\,\,
log(\difftransp{\exprind{1}})*\difftranst{\exprind{2}}) * (\difftransp{\exprind{1}}**\difftransp{\exprind{2}})
)
\\
(D-Sin) & \difftrans{sin(\exprind{1})} &=&
\adpair{
sin(\difftransp{\exprind{1}})
}{
\difftranst{\exprind{1}} * cos(\difftransp{\exprind{1}})
}
\\
(D-Cos) & \difftrans{cos(\exprind{1})} &=&
\adpair{
cos(\difftransp{\exprind{1}})
}{
-\difftranst{\exprind{1}} * sin(\difftransp{\exprind{1}})
}
\\
(D-Tan) & \difftrans{tan(\exprind{1})} &=&
\adpair{
tan(\difftransp{\exprind{1}})
}{
\difftranst{\exprind{1}} / (cos(\difftransp{\exprind{1}}) ** 2)
}
\\
(D-Log) & \difftrans{log(\exprind{1})} &=&
\adpair{
log(\difftransp{\exprind{1}})
}{
\difftranst{\exprind{1}} / \difftransp{\exprind{1}}
}
\\
(D-Exp) & \difftrans{exp(\exprind{1})} &=&
\adpair{
exp(\difftransp{\exprind{1}})
}{
\difftranst{\exprind{1}} * exp(\difftransp{\exprind{1}})
}
\\
\\
(DT-Fun) &
\difftranstype{\typefunone{\typeind{1}}{\typeind{2}}} &=&
\typefunone{\difftranstype{\typeind{1}}}{\difftranstype{\typeind{2}}} \\
(DT-Exp) &
\difftranstype{\typenum} &=&
\typenum{} $\times$ \typenum \\
(DT-Arr) &
\difftranstype{\typearray{\typemat{}}} &=&
\typearray{\difftranstype{\typemat{}}}
\\
(DT-Pair) &
\difftranstype{\typepair{$\typemat{}_1$}{$\typemat{}_2$}} &=&
\typepair{\difftranstype{$\typemat{}_1$}}{\difftranstype{$\typemat{}_2$}}
\\
\end{tabular}
\caption{Automatic Differentiation Rules for \lafsharp{} Expressions.}
\label{fig:diff_trans}
\end{figure*}


%% file: perturb.tex

\subsection{Perturbation Confusion and Nested Differentiation}
\label{sec:perturb}

In several problems such as computing the Hessian matrix, one requires to compute the differentiation of a differentiated program.
In such cases, one should be careful on dealing with tangent parts.
We demonstrate this problem in the next example.

\noindent
\textbf{Example.}
Consider the following expression:

\codespace{}

$$\frac{\partial (x \frac{\partial x + y}{\partial y})}{\partial x}$$

\codespace{}

\noindent
This expression should be evaluated to 1 at every point. 
However, an AD tool can mistakenly evaluate this expression to 2.
This is because of confusing the tangent part (perturbation) of the free variable x with the tangent of the variable y, while computing the inner derivative.
This is known as the \textit{perturbation confusion} problem in the AD literature.

We show how \system avoids this problem by using the \derivk{} macro. This expression is implemented as follows in the \fsmooth{} language:

\codespace{}

\begin{tabular}{l}
\vabs{x y}{} \\
\tab
snd (\\
\tab \tab 
\derivk{} (x * (snd (\\
\tab \tab \tab \deriv{(x + y)}{y}\\
\tab \tab
))) x
\\
\tab )
\end{tabular}

\codespace{}

\noindent
After expanding the inner \derivk{} macro, the following expression is derived:

\codespace{}

\begin{tabular}{l}
\vabs{x y}{} \\
\tab
snd (\\
\tab \tab 
\derivk{} (x * (snd (\\
\tab \tab \tab (\vabs{\diffvarprefix{x} \diffvarprefix{y}}{\adpair{\pterm{\diffvarprefix{x}} + \pterm{\diffvarprefix{y}}}{\dterm{\diffvarprefix{x}} + \dterm{\diffvarprefix{y}}}}) \adpair{x}{0} \adpair{y}{1}\\
\tab \tab
))) x
\\
\tab )
\end{tabular}

\codespace{}

\noindent
After partially evaluating the inner expression we have:

\codespace{}

\begin{tabular}{l}
\vabs{x y}{} \\
\tab
snd (\\
\tab \tab 
\derivk{} x x
\\
\tab )
\end{tabular}

\codespace{}

\noindent
Expanding this \derivk{} macro results in the following expression:

\codespace{}

\begin{tabular}{l}
\vabs{x y}{} \\
\tab
snd (\\
\tab \tab 
(\vabs{\diffvarprefix{x}}{} \diffvarprefix{x}) \adpair{x}{1}
\\
\tab )
\end{tabular}

\codespace{}

\noindent
This expression equivalent to the following expression after partial evaluation:

\codespace{}

\begin{tabular}{l}
\vabs{x y}{} \\
\tab
1
\end{tabular}

\demo

Correctly handling the perturbation confusion problem is an important feature, enabling \system to efficiently handle 
nested differentiation constructs such as computing the Hessian matrix. 
We plan to investigate the support for the Hessian matrix for the future.

%% file: efficient.tex
\section{Efficient Differentiation}
\label{sec:fsmooth_trans}

In this section, we show how \system achieves efficient differentiable programming. 
First, we show several transformation rules applicable on \fsmooth{} expressions.  
Then, we show how we generate C code from \fsmooth{} expressions for a more efficient memory management.
\subsection{Transformation Rules}
There are various algebraic identities that one can define for \fsmooth{}. 
Based on these identities, differentiated programs can be heavily optimized.
Figure~\ref{fig:fsmooth_opts} shows a set of optimizations defined for \fsmooth{}.

\input{form/fsmooth-opt}

There are various optimizations defined for scalar operations based on the ring structure of addition and multiplication, which are shown in Figure~\ref{fig:fsmooth_opt_ring}.
Note that other ring-based algebraic identities, such as associativity and commutativity, do not appear directly in the list of rules that \system applies. 
This is because they do not necessarily improve the performance, unless they are combined with other rewrite rules. 

As \fsmooth{} is based on $\lambda$-calculus, all partial evaluation rules for this calculus come for free.
Furthermore, the optimizations defined in the literature for let-binding can also be used.
Finally, partial evaluation rules for conditionals are also available. Figure~\ref{fig:fsmooth_opt_lambda} shows this set of rules.

As the vector constructs of \fsmooth{} are based on pull arrays, one can use the pull-array fusion rules for removing unnecessary intermediate vectors and matrices. The two fusion rules for pull-arrays are shown in Figure~\ref{fig:fsmooth_opt_fusion}.

In addition, many intermediate tuples resulting from the dual number technique of AD can be removed by using partial evaluation. Figure~\ref{fig:fsmooth_opt_tuple} shows the partial evaluation rules for removing the intermediate tuples which are followed by a projection.

Partially evaluating the tuples across the boundary of a loop  requires a sophisticated analysis of the body of the loop.
To simplify this task, we perform loop fission for the loops that return a tuple of values. 
This is possible only when different elements of the tuple are computed independently in different iterations of the loop.
Figure~\ref{fig:fsmooth_opt_fission} shows how loop fission turns an iteration creating a pair of elements into a pair of two iterations constructing independently the elements of that pair.
After performing this optimization, if we are interested only in a particular element of the result tuple, other loops corresponding to irrelevant elements are removed by partial evaluation.

Based on these rewrite rules, \system derives well-known matrix calculus rules, without requiring to add a rewrite rule in the level of matrices (i.e., \ladsl{}). 
However, as we will see, the order in which these rewrite rules should be applied can become tricky and for the moment are defined manually in \system. 
We leave an automatic way of inferring a good sequence of rewrite rules for the future work.

The next example, shows how \system can derive a well-known matrix identity by using a sequence of transformation rules defined in this section.

\noindent \textbf{Example 4.} Based on matrix calculus derivative rules, it is known that $\frac{\partial v_1\cdot v_2}{\partial v_1}=v_2$, where $\cdot$ is the vector dot product operator. 
We would like to show how \system{} can deduce the same algebraic identity. The differentiation of dot product of two vectors is represented as follows:

\codespace{}
\begin{tabular}{l}
\vabs{v1 v2}{}\\
\tabt{}vectorMap (\derivk{} (vectorDot v1 v2) v1) \vsndk{}
\end{tabular}

\codespace{}
\noindent This expression is expanded as follows:

\codespace{}
\begin{tabular}{l}
\vabs{v1 v2}{}\\
\tabt{}vectorMap (\\
\tabt{}\tabt{}\vbuildk{}\,(\vlengthk{}\,v1) (\vabs{i}{}\\
\tabt{}\tabt{}\tabt{}\difftrans{\vabs{v1 v2}{}vectorDot v1 v2} \\
\tabt{}\tabt{}\tabt{}\tabt{}(vectorZip v1 (vectorHot (\vlengthk{}\,v1) i)) \\
\tabt{}\tabt{}\tabt{}\tabt{}(vectorZip v2 (vectorZeros (\vlengthk{}\,v2))))\\
\tabt{}) \vsndk{}
\end{tabular}

\codespace{}
\noindent After inlining the definition of vectorMap (cf. Figure~\ref{fig:ladsl_ops}) and applying the fusion rule (cf. Figure~\ref{fig:fsmooth_opt_fusion}), the following program is produced:

\codespace{}
\begin{tabular}{l}
\vabs{v1 v2}{}\\
\tabt{}\vbuildk{}\,(\vlengthk{}\,v1) (\vabs{i}{}\\
\tabt{}\tabt{}\vsndk{} (\difftrans{\vabs{v1 v2}{}vectorDot v1 v2} \\
\tabt{}\tabt{}\tabt{}\tabt{}\tabt{}(vectorZip v1 (vectorHot (\vlengthk{}\,v1) i)) \\
\tabt{}\tabt{}\tabt{}\tabt{}\tabt{}(vectorZip v2 (vectorZeros (\vlengthk{}\,v2)))))\\
\end{tabular}

\codespace{}
\noindent After inlining the definition of vectorDot, vectorZip, vectorHot, and vectorZeros, and again applying the fusion rule, we have:

\codespace{}
\begin{tabular}{l}
\vabs{v1 v2}{}\\
\tabt{}\vbuildk{}\,(\vlengthk{}\,v1) (\vabs{i}{}\\
\tabt{}\tabt{}\vsndk{} (\difftrans{\vabs{v1 v2}{}\viteratek{} (\vabs{s j}{} s\vconst{+}\vget{v1}{j}\vconst{*}\vget{v2}{j}) 0 (\vlength{v1})} \\
\tabt{}\tabt{}\tabt{}\tabt{}\tabt{}(\vbuildk{} (\vlength{v1}) (\vabs{j}{}\adpair{\vget{v1}{j}}{\iif{}(i=j) \then{} 1  \elsee{} 0})) \\
\tabt{}\tabt{}\tabt{}\tabt{}\tabt{}(\vbuildk{} (\vlength{v2}) (\vabs{j}{}\adpair{\vget{v2}{j}}{0}))))\\
\end{tabular}

\codespace{}
\noindent After applying AD transformation rules (cf. Figure~\ref{fig:diff_trans}), and partial evaluation rules (cf. Figure~\ref{fig:fsmooth_opts}) the following program is derived:

\codespace{}
\begin{tabular}{l}
\vabs{v1 v2}{}\\
\tabt{}\vbuildk{}\,(\vlengthk{}\,v1) (\vabs{i}{}\\
\tabt{}\tabt{}\vsndk{} (\vabs{\diffvarprefix{v1} \diffvarprefix{v2}}{}\viteratek{} (\vabs{s j}{} \\
\tabt\tabt\tabt\tabt\tabt\tabt
(
(\vfstk{} s) \vconst{+} (\vfstk{} \vget{\diffvarprefix{v1}}{j}) \vconst{*} (\vfstk{} \vget{\diffvarprefix{v2}}{j})
,\\
\tabt\tabt\tabt\tabt\tabt\tabt\tabt
(\vsndk{} s) \vconst{+} 
(\vfstk{} \vget{\diffvarprefix{v1}}{j}) \vconst{*} (\vsndk{} \vget{\diffvarprefix{v2}}{j}) \vconst{+}
(\vsndk{} \vget{\diffvarprefix{v1}}{j}) \vconst{*} (\vfstk{} \vget{\diffvarprefix{v2}}{j})
)
\\
\tabt\tabt\tabt\tabt\tabt\tabt{})
\adpair{0}{0} (\vlength{\diffvarprefix{v1}})) \\
\tabt{}\tabt{}\tabt{}\tabt{}\tabt{}(\vbuildk{} (\vlength{v1}) (\vabs{j}{}\adpair{\vget{v1}{j}}{\iif{}(i=j) \then{} 1  \elsee{} 0})) \\
\tabt{}\tabt{}\tabt{}\tabt{}\tabt{}(\vbuildk{} (\vlength{v2}) (\vabs{j}{}\adpair{\vget{v2}{j}}{0}))))\\
\end{tabular}

\codespace{}
\noindent After further applying $\beta$-reduction (cf. Figure~\ref{fig:fsmooth_opt_lambda}), tuple partial evaluation (cf. Figure~\ref{fig:fsmooth_opt_lambda}), and loop fusion the following program is generated:

\codespace{}
\begin{tabular}{l}
\vabs{v1 v2}{}\\
\tabt{}\vbuildk{}\,(\vlengthk{}\,v1) (\vabs{i}{}\\
\tabt{}\tabt{}\vsndk{} (\viteratek{} (\vabs{s j}{} \\
\tabt\tabt\tabt\tabt\tabt\tabt
(
(\vfstk{} s) \vconst{+} \vget{v1}{j} \vconst{*} \vget{v2}{j}
,\\
\tabt\tabt\tabt\tabt\tabt\tabt\tabt
(\vsndk{} s) \vconst{+} 
\vget{v1}{j} \vconst{*} 0 \vconst{+}
(\iif{} (i=j) \then{} 1 \elsee{} 0) \vconst{*} \vget{v2}{j}
)
\\
\tabt\tabt\tabt\tabt\tabt\tabt{})
\adpair{0}{0} (\vlength{v1})) \\
\end{tabular}

\codespace{}
\noindent Now we apply loop fission (cf. Figure~\ref{fig:fsmooth_opt_fission}), conditional rules (cf. Figure~\ref{fig:fsmooth_opt_if}), and several other simplification rules:

\codespace{}
\begin{tabular}{l}
\vabs{v1 v2}{}\\
\tabt{}\vbuildk{}\,(\vlengthk{}\,v1) (\vabs{i}{}\\
\tabt{}\tabt{}\vsndk{} (\\
\tabt{}\tabt{}\tabt{} \viteratek{} (\vabs{s j}{}
s \vconst{+} \vget{v1}{j} \vconst{*} \vget{v2}{j}) 0 (\vlength{v1})
,\\
\tabt{}\tabt{}\tabt{} \viteratek{} (\vabs{s j}{}
\iif{} (i=j) \then{} s \vconst{+} \vget{v2}{j} \elsee{} s) 0 (\vlength{v1})\\
\tabt\tabt{}) )
\end{tabular}

\codespace{}
\noindent Note that applying the loop fission rule, does not necessarily improve the performance; 
it is only after performing tuple partial evaluation rules that the iteration responsible for the original computation is removed and the performance is improved. 
Thus, the strategy for applying rewrite rules can become tricky, and for this example, we manually specify the sequence of transformations that should be applied.
After applying the partial evaluation rule, the following program is derived:

\codespace{}
\begin{tabular}{l}
\vabs{v1 v2}{}\\
\tabt{}\vbuildk{}\,(\vlengthk{}\,v1) (\vabs{i}{}\\
\tabt{}\tabt{}(\vifoldk{}\,(\vabs{s j}{}\\
\tabt{}\tabt{}\tabt{}\iif (i = j) \then\\
\tabt{}\tabt{}\tabt{}\tabt{}(s + \vget{v2}{j})\\
\tabt{}\tabt{}\tabt{}\elsee\\
\tabt{}\tabt{}\tabt{}\tabt{}s) 0 (\vlengthk{}\,v1)))
\end{tabular}

\codespace{}
\noindent By using the optimization that turns single access iterations into a single statement (cf. Figure~\ref{fig:fsmooth_opt_iteration}), \system produces the following program:

\codespace{}
\begin{tabular}{l}
\vabs{v1 v2}{}\\
\tabt{}\vbuildk{}\,(\vlengthk{}\,v1) (\vabs{i}{}
\vget{v2}{i})
\end{tabular}

\codespace{}
\noindent This program is equivalent to $v2$ if the size of the two input vectors are the same (i.e., \vlengthk{} $v1$ = \vlengthk{} $v2$). 
Otherwise, the input program is ill-formed.

\demo

\noindent Next, we show the power of \system in deriving a matrix calculus identity for gradient of matrices.

\noindent \textbf{Example 5.} By using the same set of optimizations, \system can deduce the identity $\frac{\partial tr\big(M\big)}{\partial M}=I$. First, we start from the representation of this gradient in \fsmooth{}:

\codespace{}
\begin{tabular}{l}
\vabs{m}{}\\
\tabt{}\vbuildk{}\,(\vlengthk{}\,m) (\vabs{i}{}\\
\tabt{}\tabt{}\vbuildk{}\,(\vlengthk{}\,\vget{m}{0}) (\vabs{j}{}\\
\tabt{}\tabt{}\tabt{}\vsndk{}\,(\difftrans{matrixTrace} (matrixZip m (matrixHot (\vlengthk{}\,m) (\vlengthk{}\,\vget{m}{0}) i j)))))
\end{tabular}

\codespace{}
\noindent After applying the AD transformations and the optimizations presented in this section, the following program is produced:

\codespace{}
\begin{tabular}{l}
\vabs{m}{}\\
\tabt{}\vbuildk{}\,(\vlengthk{}\,m) (\vabs{i}{}\\
\tabt{}\tabt{}\vbuildk{}\,(\vlengthk{}\,\vget{m}{0}) (\vabs{j}{}\\
\tabt{}\tabt{}\tabt{}\iif{} (j = i) \then{} 1 \elsee{} 0))
\end{tabular}

\codespace{}
\noindent If the rows and columns of the input matrix are equal, this program represents the identity matrix with the same dimensions as the input matrix.

\demo

\noindent Similarly, \system automatically discovers the following identity if $A$ is independent of M:  $\frac{\partial tr\big(MA\big)}{\partial M}=A^T$.
Now we return to the example shown in the beginning of this paper.

\codespace{}
\noindent \textbf{Example 1 (Continued).} If we have a matrix $M$ and two vectors $u$ and $v$ (which are represented as row matrices and are independent of $M$), using matrix calculus one can prove that $\frac{\partial \big(uMv^T\big)}{\partial M}=u^Tv$. 
First, we start by a partially inlined representation of this program in \fsmooth:



\codespace{}
\begin{tabular}{l}
\lett{} f = \vabs{u M v}{}\\
\tabt{}\lett{} m =\\
\tabt \tabt matrixMult \\
\tabt \tabt \tabt (\vbuildk{}\,1 (\vabs{i}{} u)) \\
\tabt \tabt \tabt (matrixMult M \\
\tabt \tabt \tabt \tabt (matrixTranspose (\vbuildk{}\,1 (\vabs{i}{} v))))\\
\tabt{}\vget{\vget{m}{0}}{0}\\
\vabs{u M v}{}\\
\tabt{}(\vbuildk{}\,(\vlengthk{}\,M) (\vabs{i}{}\\
\tabt{}\tabt{}(\vbuildk{}\,(\vlengthk{}\,\vget{M}{0}) (\vabs{j}{}\\
\tabt{}\tabt{}\tabt{}(\vsndk{}\,(\difftrans{f}\\
\tabt{}\tabt{}\tabt{}\tabt{}(vectorZip v (vectorZeros (\vlengthk{}\,v)))\\
\tabt{}\tabt{}\tabt{}\tabt{}(matrixZip M (matrixHot (\vlengthk{}\,M) (\vlengthk{}\,\vget{M}{0}) i j))\\
\tabt{}\tabt{}\tabt{}\tabt{}(vectorZip v (vectorZeros (\vlengthk{}\,v)))))))))
\end{tabular}

\codespace{}
\noindent Note that the function f is returning the only scalar element of the 1-by-1 matrix $uMv^T$.
After performing loop fusion, loop fission and partial evaluation the following program is derived:

\codespace{}
\begin{tabular}{l}
\vabs{u M v}{}\\
\tabt{}\vbuildk{}\,(\vlengthk{}\,M) (\vabs{i}{}\\
\tabt{}\tabt{}\vbuildk{}\,(\vlengthk{}\,\vget{M}{0}) (\vabs{j}{}\\
\tabt{}\tabt{}\tabt{}\vget{u}{i} \vconst{*} \vget{v}{j}))
\end{tabular}

\codespace{}
\noindent This program is equivalent to $u^Tv$ if the input program is well formed, i.e., the number of rows and columns of $M$ are the same as the length of $u$ and $v$, respectively.

\demo







\subsection{Code Generation}
After applying the optimizations mentioned in the previous section,
one can further improve the efficiency by generating programs in a low-level language with manual memory management. 
This way, the overhead of garbage collection can be removed. 
Furthermore, by using stack-discipline memory management techniques such as Destination-Passing Style (DPS)~\cite{dps_fhpc}, one can benefit from efficient bump memory allocation instead of using the expensive \texttt{malloc} and \texttt{free} calls.

\codespace{}
\noindent 
\textbf{Example 1 (Continued).} The generated C code for the optimized differentiated program is as follows:

\codespace{}

\begin{lstlisting}[language=C, numbers=none]
matrix uMv_d(storage s, vector u, matrix M, vector v) {
  matrix res = (matrix)s;
  for(int r = 0; r < M->rows; r++) {
    for(int c = 0; c < M->cols; c++) {
      res->elems[r][c] = u->elems[r] * v->elems[c];
    }
  }
  return res;
}
\end{lstlisting}

\codespace{}

\noindent The parameter \texttt{s} is the storage area allocated for storing the result matrix.

\demo

\noindent Up to now, we have only seen the cases where only the derivative part of the program was of interest. 
If we are interested in the original part of the program as well (e.g., the intermediate vectors cannot be fused), we need to store both the original and derivative parts.
In such cases, the differentiated vectors, which are represented as arrays of tuples, can be transformed into a more efficient data layout.
The well-known array of structs (AoS) to struct of arrays (SoA) transformation represents differentiated vectors as a tuple of two numeric arrays. Further partial evaluation can remove the unnecessary decoupled numeric arrays.

%% file: form/fsmooth-opt.tex
\begin{figure}
\begin{minipage}{0.45\columnwidth}
\centering
\begin{tabular}{l r c l}
\expr{} \vconst{+} 0 = 0 \vconst{+} \expr{} 
&\evalsto& 
\expr{} \\
\expr{} \vconst{*} 1 = 1 \vconst{*} \expr{} 
&\evalsto& 
\expr{} \\
\expr{} \vconst{*} 0 = 0 \vconst{*} \expr{} 
&\evalsto& 
0 \\
\expr{} \vconst{+} -\expr{} = \expr{} \vconst{-} \expr{} 
&\evalsto& 
0 \\
\exprind{0} \vconst{*} \exprind{1} \vconst{+}
\exprind{0} \vconst{*} \exprind{2}
&\evalsto& 
\exprind{0} \vconst{*} (\exprind{1} \vconst{+} \exprind{2}) \\
\end{tabular}
\subcaption{Ring-Structure Rules}
\label{fig:fsmooth_opt_ring}
\end{minipage}
\begin{minipage}{0.45\columnwidth}
\centering
\begin{tabular}{l r l l}
(\vabs{x}{} \exprind{0}) \exprind{1}
&\evalsto& 
\exprind{0}[x $\mapsto$ \exprind{1}] \\
let x = \exprind{0} in \exprind{1}
&\evalsto& 
\exprind{1}[x $\mapsto$ \exprind{0}] \\
let x = & & let y = \exprind{0} in\\
\tab let y = \exprind{0} in \exprind{1} &\evalsto& let x = \exprind{1} \\
in \exprind{2}
& & 
in \exprind{2} \\
f(let x = \exprind{0} in \exprind{1})
&\evalsto& 
let x = \exprind{0} in f(\exprind{1}) \\
\end{tabular}
\subcaption{$\lambda$-Calculus Rules}
\label{fig:fsmooth_opt_lambda}
\end{minipage}
\begin{minipage}{0.45\columnwidth}
\centering
\begin{tabular}{l r c l}
\vget{(\vbuild{\exprind{0}}{\exprind{1}})}{\exprind{2}} 
&\evalsto& 
\exprind{1}\  \exprind{2} \\
\vlength{(\vbuild{\exprind{0}}{\exprind{1}})} 
&\evalsto&
\exprind{0}
\end{tabular}
\subcaption{Fusion Rules}
\label{fig:fsmooth_opt_fusion}
\end{minipage}
\begin{minipage}{0.45\columnwidth}
\centering
\begin{tabular}{l r c l}
\vfstk{} \adpair{\exprind{0}}{\exprind{1}}
&\evalsto& 
\exprind{0} \\
\vsndk{} \adpair{\exprind{0}}{\exprind{1}}
&\evalsto& 
\exprind{1} \\
\end{tabular}
\subcaption{Tuple Partial Evaluation Rules}
\label{fig:fsmooth_opt_tuple}
\end{minipage}
\begin{minipage}{0.95\columnwidth}
\centering
\begin{tabular}{l r c}
\vifoldk{} f z 0
&\evalsto& 
z \\
\vifoldk{} (\vabs{a i}{} a) z n
&\evalsto& 
z \\
\vifoldk{} f z n
&\evalsto& 
\vifoldk{} (\vabs{a i}{} f a (i+1)) (f z 0) (n - 1) \\
\vifoldk{} (\vabs{a i}{}
\\
\tab \iif(i = j) \then{} f a i \elsee{} a) z n
&\evalsto& 
f z j
\\
\end{tabular}
\subcaption{Iteration Rules}
\label{fig:fsmooth_opt_iteration}
\end{minipage}
\hfill
\begin{minipage}{0.95\columnwidth}
\centering
\begin{tabular}{l r c}
\iif{} true \then{} \exprind{1} \elsee{} \exprind{2}
&\evalsto&  
\exprind{1}
\\
\iif{} false \then{} \exprind{1} \elsee{} \exprind{2}
&\evalsto&  
\exprind{2}
\\
\iif{} \exprind{0} \then{} \exprind{1} \elsee{} \exprind{1}
&\evalsto&  
\exprind{1}
\\
f (\iif{} \exprind{0} \then{} \exprind{1} \elsee{} \exprind{2})
&\evalsto&  
\iif{} \exprind{0} \then{} f (\exprind{1}) \elsee{} f (\exprind{2})
\\
\iif{} \exprind{0} \then{} \exprind{1} \elsee{} \exprind{2}
&\evalsto&  
\iif{} \exprind{0} \then{} \exprind{1}[\exprind{0} $\mapsto$ true] \elsee{} \exprind{2}[\exprind{0} $\mapsto$ false]
\\
\end{tabular}
\subcaption{Conditional Rules}
\label{fig:fsmooth_opt_if}
\end{minipage}
\begin{minipage}{0.95\columnwidth}
\centering
\setlength{\tabcolsep}{1pt}
\begin{tabular}{l r l}
\vifoldk{} (\vabs{a i}{}
\adpair{f (\vfstk{} a) i}{g (\vsndk{} a) i}
) \adpair{z1}{z2} n
&\evalsto& 
\adpair{\vifoldk{} f z1 n}{\vifoldk{} g z2 n}
\\
\end{tabular}
\subcaption{Loop Fission}
\label{fig:fsmooth_opt_fission}
\end{minipage}
\caption{Optimizations for \fsmooth{}.}
\label{fig:fsmooth_opts}
\end{figure}

%% file: exp.tex
\section{Experimental Results}
\label{sec:exp}
In this section, we show how \system performs in practice.
We show the performance of the differentiated code for two real-world machine learning and computer vision applications.

\noindent \textbf{Experimental Setup.} We have performed the experiments using an iMac machine equipped with
an Intel Core i5 CPU running at 2.7GHz, 32GB of DDR3 RAM at
1333Mhz.
The operating system is OS X 10.13.1. We use CLang 900.0.39.2 for compiling the generated C code, and Python 2.7.12 for running the Python code.

\noindent \textbf{Micro Benchmarks} which consist of the following vector expressions: 1) gradient of dot product of two vectors with respect to the first vector (which is a Jacobian matrix with a single row), 2) gradient of the maximum value of a vector with respect to the input vector (which is a Jacobian matrix with a single row), 3) gradient of addition of two vectors with respect to the first vector (which is a Jacobian matrix), and 4) gradient of the multiplication of a vector with a scalar value with respect to the scalar value (which is a Jacobian matrix with a single column). 

\begin{figure}[t!]
        \centering
    	\includegraphics[width=0.5\textwidth]{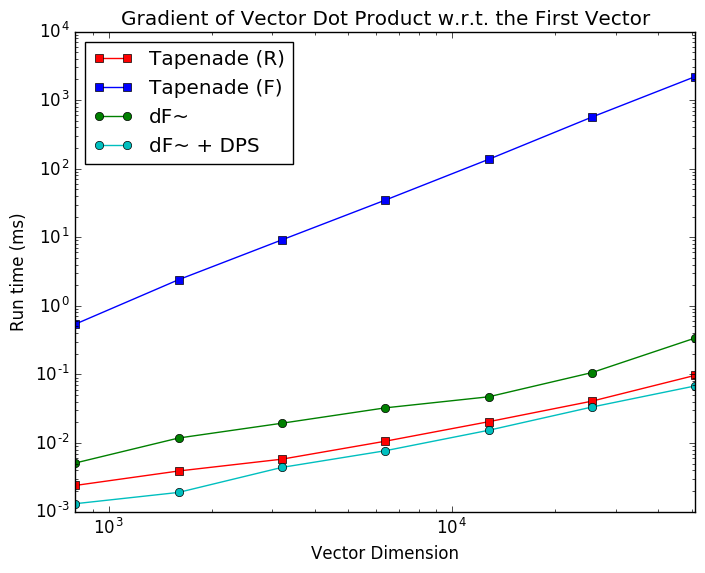}~\includegraphics[width=0.5\textwidth]{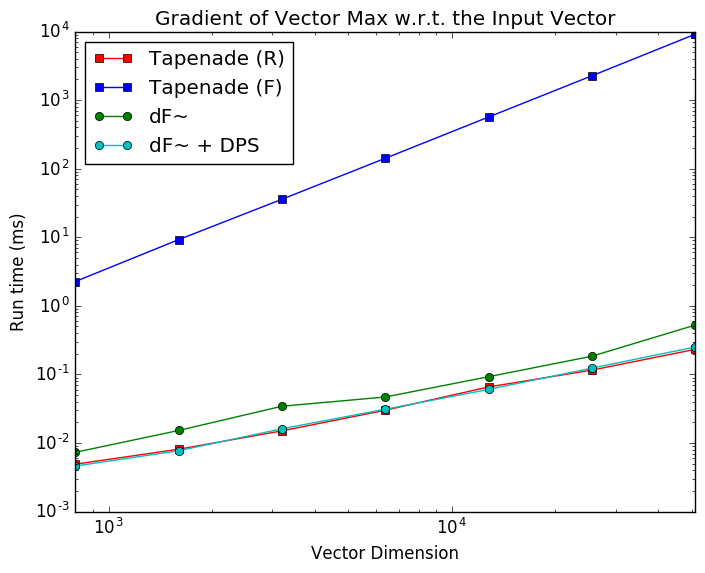}\\
        \includegraphics[width=0.5\textwidth]{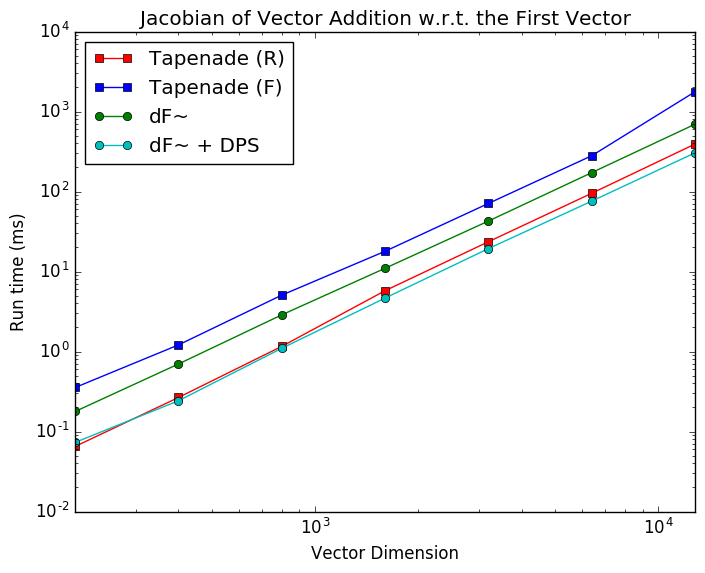}~\includegraphics[width=0.5\textwidth]{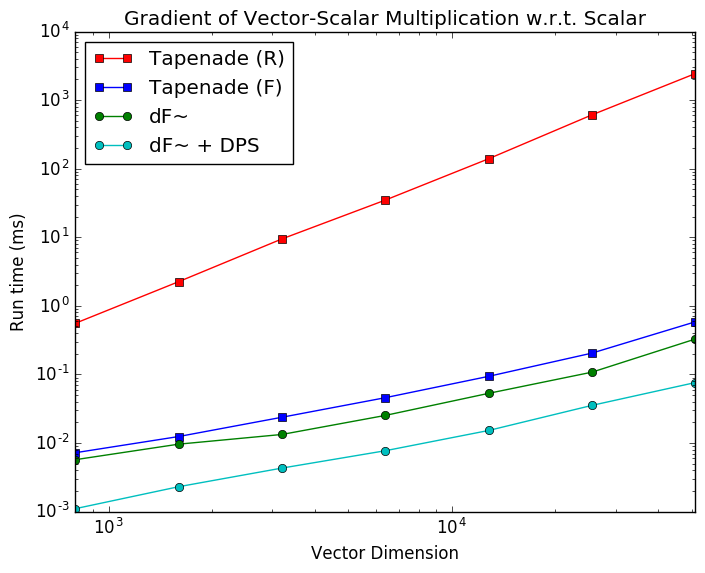}
        \caption{Performance results for Micro Benchmarks.}
        \label{micro_perf}
\end{figure}

Figure~\ref{micro_perf} shows the performance results for the mentioned micro benchmarks for \system and both forward and reverse-mode of Tapenade. 
In all cases, \system{} outperforms or performs as good as both forward and reverse-mode of Tapenade. 
The performance is improved further when the generated C code uses Destination-Passing Style (DPS)~\cite{dps_fhpc} for stack-discipline memory management. 

As in the first two cases the Jacobian matrix is a row vector, reverse-mode AD computes the whole Jacobian matrix in a single backward pass. However, forward-mode AD needs to iterate over each column to compute the corresponding derivative value.
For the case of the addition of two vectors, as the Jacobian matrix is a square matrix, reverse-mode AD and forward-mode AD show comparable performance.
Finally, for the last case, as the Jacobian matrix is a column vector, the forward mode AD computes the whole Jacobian matrix in a single forward pass.
However, the reverse mode AD requires traversing over each row to compute the corresponding partial derivative values.

\noindent \textbf{Non-Negative Matrix Factorization (NNMF)} is a useful tool which has many applications in various fields ranging from document clustering, recommendation systems, signal processing, to computer vision. 
For instance, in~\cite{liu2010distributed}, the authors study the NNMF of Web dyadic data represented as the matrix $A$. 
Dyadic data contains rich information about the interactions between the two participating sets. It is useful for a broad range of practical applications including Web search, Internet monetization, and social media content~\cite{liu2010distributed}. 
For example the (query, clicked URL) data is used in query clustering~\cite{2002:QCU:503104.503108}, query suggestions~\cite{Baeza-Yates:2004:QRU:2146449.2146527} and improving search relevance~\cite{Agichtein:2006:IWS:1148170.1148177}. 
Matrix factorization is a commonly used approach to understanding the latent structure of the observed matrix for various applications~\cite{Berry06algorithmsand, Sra06nonnegativematrix}. The authors present a probabilistic NNMF framework for a variety of Web dyadic data that conforms to different probabilistic distributions. For instance, an Exponential distribution is used to model Web lifetime dyadic data, e.g., user dwell time, and similarly the Poisson distribution is used to model count dyadic data, e.g., click counts.

\begin{figure}[t!]
        \centering
    	\includegraphics[width=0.41\textwidth]{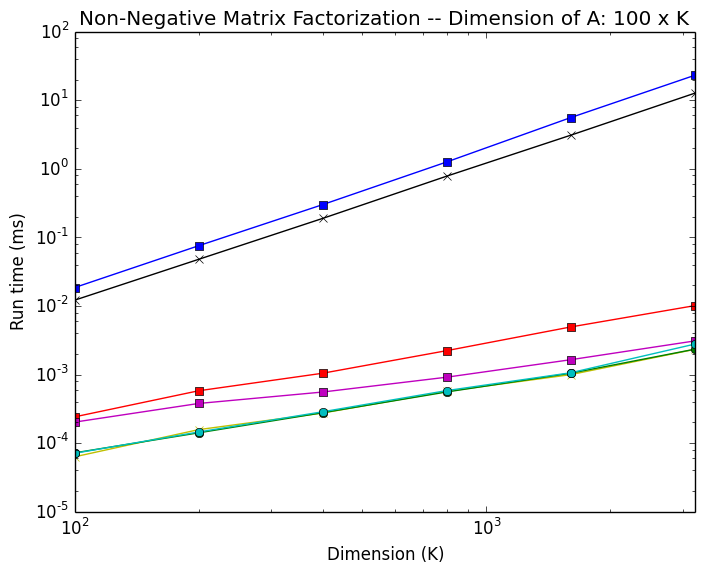}~\includegraphics[width=0.59\textwidth]{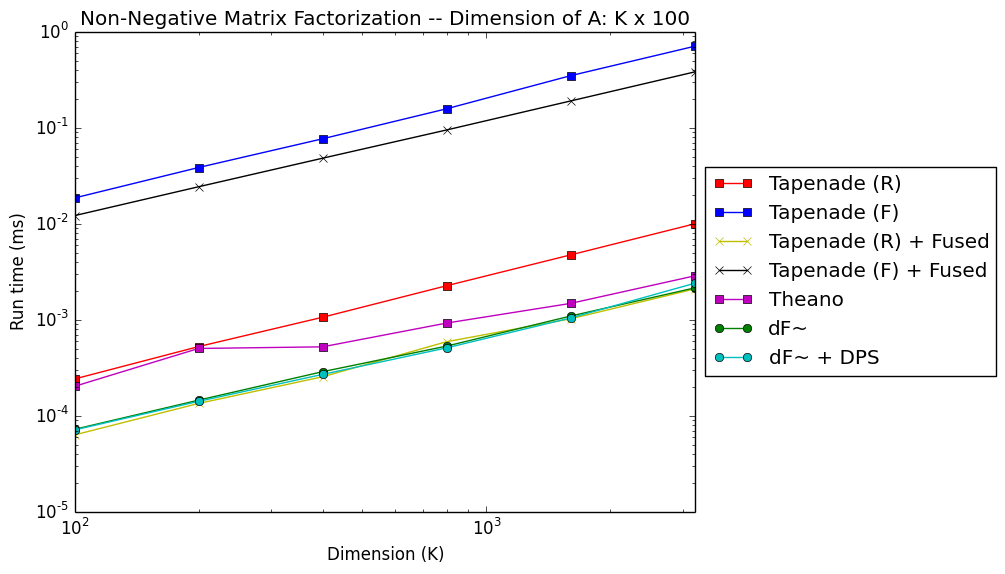}
        \caption{Performance results for NNMF.}
        \label{nmf_perf}
\end{figure}

The iterative algorithm to find $W$ and $H$ depends on the form of the assumed underlying distribution. 
In particular the update formula for gradient descent are derived by computing the gradient of the negative $\log$ of the likelihood function. 
For example, the negative $\log$ of the exponential distribution is represented as follows:

$$\mathcal{D}\big(A||\widetilde{A}\big)=\mathlarger{\Sigma}_{(i,j)}\Bigg(\log\big(\widetilde{A}_{i,j}\big) + \frac{A_{i, j}}{\widetilde{A}_{i,j}}\Bigg) ,\hspace{1cm} \widetilde{A} = WH$$

\noindent The update formulas are derived manually, and for each new distribution it is the responsibility of the user to undertake the error prone and laborious task of deriving, optimizing, and implementing the update rules. 
\system automatically derives the gradient of the negative $\log$ of the likelihood function for the exponential distribution. After performing optimizations, \system produces an expression which is equivalent to the following update formula, which is manually derived by hand in~\cite{liu2010distributed}:

$$\frac{\partial \mathcal{D}}{\partial H} = W^T\Bigg(\frac{1}{WH}-\frac{A}{\big(WH\big)^2}\Bigg)$$

\noindent Figure~\ref{nmf_perf} shows the performance results of executing the derived update rule on Tapenade, Theano, and \system.
For all the experiments, we consider factorizing the matrix $A$ into two vectors $W$ and $H$ (represented as $u$ and $v^T$, respectively).
To have a fair comparison between Tapenade and \system, we have provided both the fused and unfused versions of the likelihood function. 
We observe a 2x speed up for the forward mode, and a 5x speed up for the reverse mode, when comparing the fused version with the unfused version.
Comparing the fused version of Tapenade and \system, we observe that the reverse-mode AD of Tapenade behaves similarly to \system.
This shows that \system successfully generates efficient code for this case, which is an ideal case for the reverse-mode AD (the loss function is a scalar valued function, which should compute the gradient with respect to all elements of the input vector).
Finally, as the dimension of the vectors increases, Theano converges to the same performance as \system and reverse-mode AD of Tapenade. 
This is thanks to the fact that the overhead of invoking C functions from Python becomes negligible as the size of the vector increases.

\noindent \textbf{The Gaussian Mixture Model (GMM)} is a statistical method used for various machine learning tasks such as unsupervised and semi-supervised learning, as well as computer vision applications such as image background modelling and image denoising. 

Here we focus on computing the gradient of one of function used in GMM: the Log-Sum-Exp (LSE) of a vector is useful in various machine learning algorithms such as GMM~\cite{nielsen2016guaranteed,huang2008new}.
Intuitively, if the multiplication operation in the linear domain is transformed into addition in the log domain, the addition operation is transformed into LSE in the log domain.
This expression is computed as follows.

$$LSE\big(x_1, ..., x_n\big) = x_{max} + log\big(\Sigma_{i=1}^n(e^{x_i - x_{max}})\big)$$

\begin{figure}[t!]
        \centering
    	\includegraphics[width=0.65\textwidth]{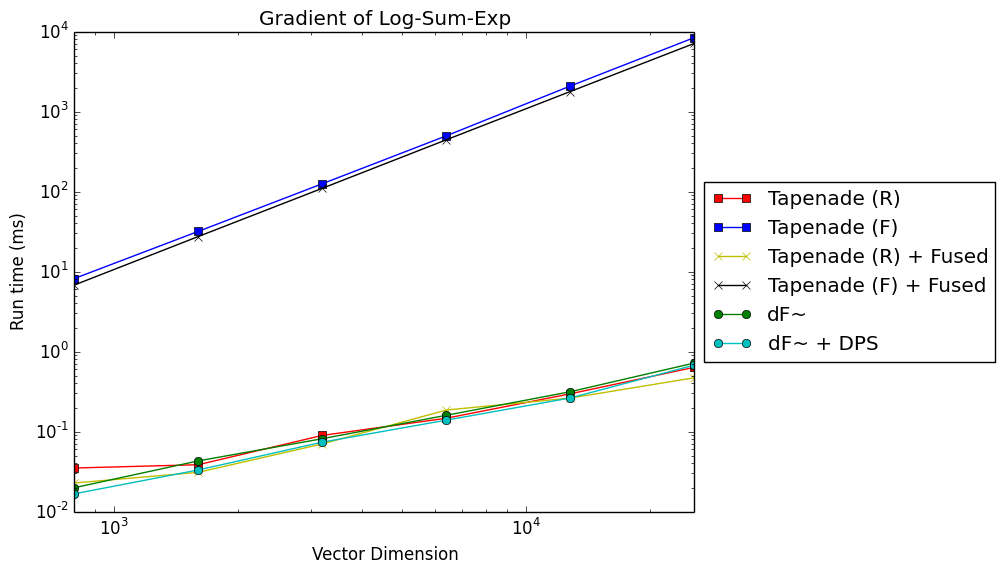}
        \caption{Performance results for log-sum-exp used in GMM.}
        \label{lse_perf}
\end{figure}

\noindent Figure~\ref{lse_perf} shows the performance results for the gradient of this function with respect to its input vector. 
Applying fusion improves the performance of the differentiated programs by 25\%.
Comparing the fused versions of the programs, \system{} outperforms the forward-mode AD of Tapenade from 2 to 4 orders of magnitude. This gap increases quadratically with the size of the input vector.
However, \system{} shows a similar performance to the fused reverse-mode AD of Tapenade.

\noindent \textbf{Bundle Adjustment} \cite{triggs1999bundle,agarwal2010bundle,zach2014robust} is a computer vision problem, where the goal is to optimize several parameters in order to have an accurate estimate of the projection of a 3D point by a camera. 
This is achieved by minimizing an objective function representing the reprojection error. 

For the experiments, we compute the Jacobian matrix of the Project function in Bundle Adjustment. 
For a 3D point $X\in \mathbb{R}^3$ and a camera with rotation parameter $r \in \mathbb{R}^3$, center position $C \in \mathbb{R}^3$, focal index $f \in \mathbb{R}$, principal point $x_0 \in \mathbb{R}^2$, and radical distortion $k \in \mathbb{R}^2$, the Project function is computes the projected point as follows:
\\ \\
\begin{tabular}{r @{\hskip3pt} c @{\hskip3pt} l}
\setlength\tabcolsep{.5pt}
$\text{project}\big(r, C, f, x_0, k, X\big)$
& = &
$\text{distort}\big(k, \text{p2e}\big(\text{rodrigues}\big(r, X - C\big)\big)\big)f + x_0$
\\
$\text{distort}\big(k, x\big)$
&=&
$x\big(1 + k_1||x||^2+k_2||x||^4\big)$
\\
$\text{p2e}\big(X\big) $
&=&
$X_{1..2}/X_3$
\\
$\text{rodrigues}\big(r, X\big)$
&=&
$X cos \theta + \big(v \times X\big)sin \theta + v\big(v^TX\big)\big(1 - cos \theta\big), \theta=||r||, v=\frac{r}{||r||}$
\end{tabular} 
\\ \\

\begin{figure}[t!]
        \centering
    	\includegraphics[width=0.5\textwidth]{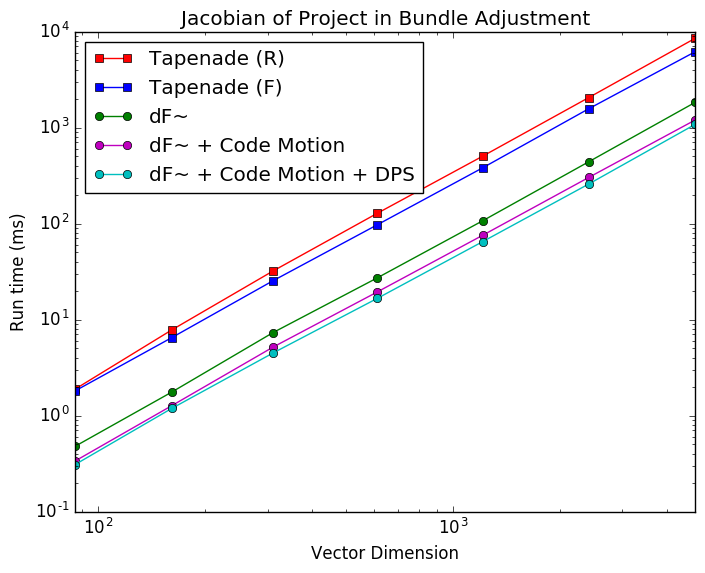}
        \caption{Performance results for Project in Bundle Adjustment.}
        \label{ba_perf}
\end{figure}

\noindent Consider having $N$ 3D points and one particular camera parameter (an input vector of size $3N+11$), we are interested in computing a Jacobian matrix with $3N+11$ rows and $2N$ columns.
Figure~\ref{ba_perf} shows the performance results for  computing the mentioned Jacobian matrix.
As it can be seen \system outperforms both forward  and reverse mode of Tapenade. 
This is mainly thanks to the loop transformations, such as loop-invariant code motion, happening in \system. 

%% file: related.tex
\section{Related Work}
\label{sec_related}
\noindent \textbf{Automatic Differentiation. }
There is a large body of work on automatic differentiation (AD) of imperative programming languages.
Tapenade~\cite{tapenade} performs AD for a subset of C and Fortran, whereas, ADIFOR~\cite{bischof1996adifor} performs AD for Fortran programs.
Adept~\cite{adept} and ADIC~\cite{narayanan2010adic2} perform automatic differentiation for C++ by using expression templates. 
However, as we have seen in our experimental results, an AD tool such as Tapenade misses several optimization opportunities, mainly due to their limited support for loop fusion.

ADiMat~\cite{bischof2002combining}, ADiGator~\cite{weinstein2016algorithm}, and Mad~\cite{forth2006efficient} perform AD for MATLAB programs, whereas MuPAD~\cite{hivert2004mupad} computes the derivatives using symbolic differentiation.
AutoGrad~\cite{maclaurin2015autograd} performs AD for Python programs that use NumPy library for array manipulation, whereas
Theano~\cite{bergstra2010theano} uses symbolic differentiation. Tensorflow~\cite{abadi2016tensorflow} performs source-to-source reverse-mode AD, and uses advanced heuristics to solve the memory inefficiencies.
ForwardDiff~\cite{revels2016forward} employs vector forward-mode AD~\cite{khan2015vector} for differentiating Julia programs. This system keeps a vector of derivative values in the dual number instead of only a single derivative value.
All these systems miss important optimization opportunities such as loop fusion.

DiffSharp~\cite{baydin2015diffsharp} is an AD library implemented in F\#. 
This library provides both forward-mode and reverse-mode AD techniques.
As DiffSharp is a library implementation of AD (in contrast to \system, which implements AD as source-to-source transformation rules), 
it cannot not support the simplification rules such as loop-invariant code motion, loop fusion, and partial evaluation. 
Furthermore, \system can efficiently manage memory by generating C code using DPS, whereas DiffSharp should rely on the garbage collection provided by the .NET framework for memory management. 

Stalingrad~\cite{pearlmutter2008reverse} is an optimizing compiler for a dialect of Scheme with a first-class AD operator, with the support for both forward mode and reverse mode of AD. One of the key challenges that Stalingrad addresses is perturbation confusion~\cite{siskind2005perturbation}, which occurs for computing the derivative of the functions for which the derivatives are already computed, or the cases where we need the computation of nested differentiation~\cite{pearlmutter2007lazy}. 
We have shown in Section~\ref{sec:perturb} how \system resolves the perturbation confusion problem.
One key limitation of Stalingrad is the lack of support for variable-size vectors; Stalingrad only supports a statically-known-size list of elements which are unfolded using Scheme macros.

Karczmarczuk~\cite{karczmarczuk1999functional} presents a Haskell implementation for both forward and reverse mode AD.
Elliott~\cite{elliott2009beautiful} improves this work by giving a more elegant implementation for its forward mode AD. 
These implementations lack the optimizations offered by transformation rules, such as loop fusion.

\noindent \textbf{Array Languages and Fusion.}
There are many array programming languages in the literature, 
APL~\cite{iverson1962programming} being the pioneer among them. 
There are functional array languages such as Futhark~\cite{henriksen2017futhark} and SAC~\cite{Grelck2006} with support for fusion.

In array languages fusion can be achieved by using functional arrays known as \emph{push} and \emph{pull arrays}~\cite{Svensson:2014:DPA:2636228.2636231,edsl-push,Claessen:2012:EAC:2103736.2103740}. A push-array is represented by an effectful function that, given an index and a value, will write the value into the array. 
A pull-array is represented by the length of the array and a function producing an element for a given index, similar to the \vbuildk{} construct in \fsmooth.
Similarly, functional programming languages use shortcut deforestation for fusing lists either by pulling the stream of data~\cite{Svenningsson:2002:SFA:581478.581491,Coutts07streamfusion} or pushing them~\cite{foldr-fusion-1}.

\noindent \textbf{Numerical DSLs.}
There are many DSLs for numerical workloads. These DSLs can be classified in three categories.
The first category consists of mainstream programming languages used by data analysts such as MATLAB and R. 
These languages offer many toolboxes for performing a wide range of tasks, however, from a performance point of view the focus is only on the efficient implementation of the libraries.
The second category consists of DSLs such as Lift~\cite{Steuwer:2015:GPP:2784731.2784754}, 
Opt~\cite{devito2016opt}, 
Halide~\cite{ragan2013halide}, Diderot~\cite{chiw2012diderot}, and OptiML~\cite{sujeeth2011optiml}, which
generate parallel code from their high-level programs. 
The third category is the DSLs which focus on generating efficient machine code for fixed size linear algbra problems such as Spiral~\cite{spiral} and LGen~\cite{spampinato2016basic}.
These DSLs exploit the memory hierarchy by relying on searching algorithms for making tiling and scheduling decisions.
Except the first category, for which automatic differentiation tools exist, the other DSLs do not have any support for automatic differentiation. 
Moreover, parallel code generation and efficient machine code generation are orthogonal concepts and can be added to \system in the future.

%% file: conclusion.tex
\section{Outlook and Conclusions}
In this paper we have demonstrated how to efficiently compute the derivate of a program.
The key idea behind our system is exposing all the constructs used in differentiated programs to the underlying compiler.
As a result, the compiler can apply various loop transformations such as loop-invariant code motion and loop fusion for optimizing differentiated programs.
We have shown how \system outperforms the existing AD tools on micro benchmarks and real-world machine learning and computer vision applications.

We plan to extend \system with the reverse-mode AD by employing a similar technique to the one proposed by~\cite{pearlmutter2008reverse}.
In addition, as we have seen in our examples, the strategy for applying rewrite rules can become tricky in some cases; there are some rewrite rules (e.g., loop fission) that do not necessarily improve the performance, unless they are combined with other transformation rules.
We plan to investigate the use of search strategies for automated rewriting (e.g., using Monte-Carlo tree search~\cite{de2009bandit}). 


%% file: main.bbl

\begin{thebibliography}{56}


\ifx \showCODEN    \undefined \def \showCODEN     #1{\unskip}     \fi
\ifx \showDOI      \undefined \def \showDOI       #1{#1}\fi
\ifx \showISBNx    \undefined \def \showISBNx     #1{\unskip}     \fi
\ifx \showISBNxiii \undefined \def \showISBNxiii  #1{\unskip}     \fi
\ifx \showISSN     \undefined \def \showISSN      #1{\unskip}     \fi
\ifx \showLCCN     \undefined \def \showLCCN      #1{\unskip}     \fi
\ifx \shownote     \undefined \def \shownote      #1{#1}          \fi
\ifx \showarticletitle \undefined \def \showarticletitle #1{#1}   \fi
\ifx \showURL      \undefined \def \showURL       {\relax}        \fi
\providecommand\bibfield[2]{#2}
\providecommand\bibinfo[2]{#2}
\providecommand\natexlab[1]{#1}
\providecommand\showeprint[2][]{arXiv:#2}

\bibitem[\protect\citeauthoryear{Abadi, Barham, Chen, Chen, Davis, Dean, Devin,
  Ghemawat, Irving, Isard, et~al\mbox{.}}{Abadi et~al\mbox{.}}{2016}]%
        {abadi2016tensorflow}
\bibfield{author}{\bibinfo{person}{Mart{\'\i}n Abadi}, \bibinfo{person}{Paul
  Barham}, \bibinfo{person}{Jianmin Chen}, \bibinfo{person}{Zhifeng Chen},
  \bibinfo{person}{Andy Davis}, \bibinfo{person}{Jeffrey Dean},
  \bibinfo{person}{Matthieu Devin}, \bibinfo{person}{Sanjay Ghemawat},
  \bibinfo{person}{Geoffrey Irving}, \bibinfo{person}{Michael Isard},
  {et~al\mbox{.}}} \bibinfo{year}{2016}\natexlab{}.
\newblock \showarticletitle{TensorFlow: A System for Large-Scale Machine
  Learning.}. In \bibinfo{booktitle}{\emph{OSDI}}, Vol.~\bibinfo{volume}{16}.
  \bibinfo{pages}{265--283}.
\newblock


\bibitem[\protect\citeauthoryear{Agarwal, Snavely, Seitz, and Szeliski}{Agarwal
  et~al\mbox{.}}{2010}]%
        {agarwal2010bundle}
\bibfield{author}{\bibinfo{person}{Sameer Agarwal}, \bibinfo{person}{Noah
  Snavely}, \bibinfo{person}{Steven~M Seitz}, {and} \bibinfo{person}{Richard
  Szeliski}.} \bibinfo{year}{2010}\natexlab{}.
\newblock \showarticletitle{Bundle adjustment in the large}. In
  \bibinfo{booktitle}{\emph{European conference on computer vision}}. Springer,
  \bibinfo{pages}{29--42}.
\newblock


\bibitem[\protect\citeauthoryear{Agichtein, Brill, and Dumais}{Agichtein
  et~al\mbox{.}}{2006}]%
        {Agichtein:2006:IWS:1148170.1148177}
\bibfield{author}{\bibinfo{person}{Eugene Agichtein}, \bibinfo{person}{Eric
  Brill}, {and} \bibinfo{person}{Susan Dumais}.}
  \bibinfo{year}{2006}\natexlab{}.
\newblock \showarticletitle{Improving Web Search Ranking by Incorporating User
  Behavior Information}. In \bibinfo{booktitle}{\emph{SIGIR}}.
\newblock


\bibitem[\protect\citeauthoryear{Anker and Svenningsson}{Anker and
  Svenningsson}{2013}]%
        {edsl-push}
\bibfield{author}{\bibinfo{person}{Johan Anker} {and} \bibinfo{person}{Josef
  Svenningsson}.} \bibinfo{year}{2013}\natexlab{}.
\newblock \showarticletitle{An {EDSL} approach to high performance Haskell
  programming}. In \bibinfo{booktitle}{\emph{ACM Haskell Symposium}}.
  \bibinfo{pages}{1--12}.
\newblock


\bibitem[\protect\citeauthoryear{Baeza-Yates, Hurtado, and Mendoza}{Baeza-Yates
  et~al\mbox{.}}{2004}]%
        {Baeza-Yates:2004:QRU:2146449.2146527}
\bibfield{author}{\bibinfo{person}{Ricardo Baeza-Yates},
  \bibinfo{person}{Carlos Hurtado}, {and} \bibinfo{person}{Marcelo Mendoza}.}
  \bibinfo{year}{2004}\natexlab{}.
\newblock \showarticletitle{Query Recommendation Using Query Logs in Search
  Engines}. In \bibinfo{booktitle}{\emph{EDBT}}.
\newblock


\bibitem[\protect\citeauthoryear{Baydin, Pearlmutter, Radul, and
  Siskind}{Baydin et~al\mbox{.}}{2015b}]%
        {baydin2015automatic}
\bibfield{author}{\bibinfo{person}{Atilim~Gunes Baydin},
  \bibinfo{person}{Barak~A Pearlmutter}, \bibinfo{person}{Alexey~Andreyevich
  Radul}, {and} \bibinfo{person}{Jeffrey~Mark Siskind}.}
  \bibinfo{year}{2015}\natexlab{b}.
\newblock \showarticletitle{Automatic differentiation in machine learning: a
  survey}.
\newblock \bibinfo{journal}{\emph{arXiv preprint arXiv:1502.05767}}
  (\bibinfo{year}{2015}).
\newblock


\bibitem[\protect\citeauthoryear{Baydin, Pearlmutter, and Siskind}{Baydin
  et~al\mbox{.}}{2015a}]%
        {baydin2015diffsharp}
\bibfield{author}{\bibinfo{person}{Atilim~Gunes Baydin},
  \bibinfo{person}{Barak~A Pearlmutter}, {and} \bibinfo{person}{Jeffrey~Mark
  Siskind}.} \bibinfo{year}{2015}\natexlab{a}.
\newblock \showarticletitle{Diffsharp: Automatic differentiation library}.
\newblock \bibinfo{journal}{\emph{arXiv preprint arXiv:1511.07727}}
  (\bibinfo{year}{2015}).
\newblock


\bibitem[\protect\citeauthoryear{Bergstra, Breuleux, Bastien, Lamblin, Pascanu,
  Desjardins, Turian, Warde-Farley, and Bengio}{Bergstra et~al\mbox{.}}{2010}]%
        {bergstra2010theano}
\bibfield{author}{\bibinfo{person}{James Bergstra}, \bibinfo{person}{Olivier
  Breuleux}, \bibinfo{person}{Fr{\'e}d{\'e}ric Bastien},
  \bibinfo{person}{Pascal Lamblin}, \bibinfo{person}{Razvan Pascanu},
  \bibinfo{person}{Guillaume Desjardins}, \bibinfo{person}{Joseph Turian},
  \bibinfo{person}{David Warde-Farley}, {and} \bibinfo{person}{Yoshua Bengio}.}
  \bibinfo{year}{2010}\natexlab{}.
\newblock \showarticletitle{Theano: A CPU and GPU math compiler in Python}. In
  \bibinfo{booktitle}{\emph{Proc. 9th Python in Science Conf}}.
  \bibinfo{pages}{1--7}.
\newblock


\bibitem[\protect\citeauthoryear{Berry, Browne, Langville, Pauca, and
  Plemmons}{Berry et~al\mbox{.}}{2006}]%
        {Berry06algorithmsand}
\bibfield{author}{\bibinfo{person}{Michael~W. Berry}, \bibinfo{person}{Murray
  Browne}, \bibinfo{person}{Amy~N. Langville}, \bibinfo{person}{V.~Paul Pauca},
  {and} \bibinfo{person}{Robert~J. Plemmons}.} \bibinfo{year}{2006}\natexlab{}.
\newblock \showarticletitle{Algorithms and applications for approximate
  nonnegative matrix factorization}. In \bibinfo{booktitle}{\emph{Computational
  Statistics and Data Analysis}}.
\newblock


\bibitem[\protect\citeauthoryear{Bischof, Khademi, Mauer, and Carle}{Bischof
  et~al\mbox{.}}{1996}]%
        {bischof1996adifor}
\bibfield{author}{\bibinfo{person}{Christian Bischof}, \bibinfo{person}{Peyvand
  Khademi}, \bibinfo{person}{Andrew Mauer}, {and} \bibinfo{person}{Alan
  Carle}.} \bibinfo{year}{1996}\natexlab{}.
\newblock \showarticletitle{ADIFOR 2.0: Automatic differentiation of Fortran 77
  programs}.
\newblock \bibinfo{journal}{\emph{IEEE Computational Science and Engineering}}
  \bibinfo{volume}{3}, \bibinfo{number}{3} (\bibinfo{year}{1996}),
  \bibinfo{pages}{18--32}.
\newblock


\bibitem[\protect\citeauthoryear{Bischof, Bucker, Lang, Rasch, and
  Vehreschild}{Bischof et~al\mbox{.}}{2002}]%
        {bischof2002combining}
\bibfield{author}{\bibinfo{person}{Christian~H Bischof}, \bibinfo{person}{HM
  Bucker}, \bibinfo{person}{Bruno Lang}, \bibinfo{person}{Arno Rasch}, {and}
  \bibinfo{person}{Andre Vehreschild}.} \bibinfo{year}{2002}\natexlab{}.
\newblock \showarticletitle{Combining source transformation and operator
  overloading techniques to compute derivatives for MATLAB programs}. In
  \bibinfo{booktitle}{\emph{Source Code Analysis and Manipulation, 2002.
  Proceedings. Second IEEE International Workshop on}}. IEEE,
  \bibinfo{pages}{65--72}.
\newblock


\bibitem[\protect\citeauthoryear{Chiw, Kindlmann, Reppy, Samuels, and
  Seltzer}{Chiw et~al\mbox{.}}{2012}]%
        {chiw2012diderot}
\bibfield{author}{\bibinfo{person}{Charisee Chiw}, \bibinfo{person}{Gordon
  Kindlmann}, \bibinfo{person}{John Reppy}, \bibinfo{person}{Lamont Samuels},
  {and} \bibinfo{person}{Nick Seltzer}.} \bibinfo{year}{2012}\natexlab{}.
\newblock \showarticletitle{{Diderot: A Parallel DSL for Image Analysis and
  Visualization}} \emph{(\bibinfo{series}{PLDI '12})}.
  \bibinfo{publisher}{ACM}, \bibinfo{pages}{111--120}.
\newblock


\bibitem[\protect\citeauthoryear{Claessen, Sheeran, and Svensson}{Claessen
  et~al\mbox{.}}{2012}]%
        {Claessen:2012:EAC:2103736.2103740}
\bibfield{author}{\bibinfo{person}{Koen Claessen}, \bibinfo{person}{Mary
  Sheeran}, {and} \bibinfo{person}{Bo~Joel Svensson}.}
  \bibinfo{year}{2012}\natexlab{}.
\newblock \showarticletitle{Expressive Array Constructs in an Embedded GPU
  Kernel Programming Language} \emph{(\bibinfo{series}{DAMP '12})}.
  \bibinfo{publisher}{ACM}, \bibinfo{address}{NY, USA},
  \bibinfo{pages}{21--30}.
\newblock


\bibitem[\protect\citeauthoryear{Coutts, Leshchinskiy, and Stewart}{Coutts
  et~al\mbox{.}}{[n. d.]}]%
        {Coutts07streamfusion}
\bibfield{author}{\bibinfo{person}{Duncan Coutts}, \bibinfo{person}{Roman
  Leshchinskiy}, {and} \bibinfo{person}{Don Stewart}.} \bibinfo{year}{[n.
  d.]}\natexlab{}.
\newblock \showarticletitle{{Stream Fusion. From Lists to Streams to Nothing at
  All}} \emph{(\bibinfo{series}{ICFP '07})}.
\newblock


\bibitem[\protect\citeauthoryear{De~Mesmay, Rimmel, Voronenko, and
  P{\"u}schel}{De~Mesmay et~al\mbox{.}}{2009}]%
        {de2009bandit}
\bibfield{author}{\bibinfo{person}{Fr{\'e}d{\'e}ric De~Mesmay},
  \bibinfo{person}{Arpad Rimmel}, \bibinfo{person}{Yevgen Voronenko}, {and}
  \bibinfo{person}{Markus P{\"u}schel}.} \bibinfo{year}{2009}\natexlab{}.
\newblock \showarticletitle{Bandit-based optimization on graphs with
  application to library performance tuning}. In
  \bibinfo{booktitle}{\emph{Proceedings of the 26th Annual International
  Conference on Machine Learning}}. ACM, \bibinfo{pages}{729--736}.
\newblock


\bibitem[\protect\citeauthoryear{DeVito, Mara, Zollh{\"o}fer, Bernstein,
  Ragan-Kelley, Theobalt, Hanrahan, Fisher, and Nie{\ss}ner}{DeVito
  et~al\mbox{.}}{2016}]%
        {devito2016opt}
\bibfield{author}{\bibinfo{person}{Zachary DeVito}, \bibinfo{person}{Michael
  Mara}, \bibinfo{person}{Michael Zollh{\"o}fer}, \bibinfo{person}{Gilbert
  Bernstein}, \bibinfo{person}{Jonathan Ragan-Kelley},
  \bibinfo{person}{Christian Theobalt}, \bibinfo{person}{Pat Hanrahan},
  \bibinfo{person}{Matthew Fisher}, {and} \bibinfo{person}{Matthias
  Nie{\ss}ner}.} \bibinfo{year}{2016}\natexlab{}.
\newblock \showarticletitle{{Opt: A Domain Specific Language for Non-linear
  Least Squares Optimization in Graphics and Imaging}}.
\newblock \bibinfo{journal}{\emph{arXiv preprint arXiv:1604.06525}}
  (\bibinfo{year}{2016}).
\newblock


\bibitem[\protect\citeauthoryear{Elliott}{Elliott}{2009}]%
        {elliott2009beautiful}
\bibfield{author}{\bibinfo{person}{Conal~M Elliott}.}
  \bibinfo{year}{2009}\natexlab{}.
\newblock \showarticletitle{Beautiful differentiation}. In
  \bibinfo{booktitle}{\emph{ACM Sigplan Notices}}, Vol.~\bibinfo{volume}{44}.
  ACM, \bibinfo{pages}{191--202}.
\newblock


\bibitem[\protect\citeauthoryear{Forth}{Forth}{2006}]%
        {forth2006efficient}
\bibfield{author}{\bibinfo{person}{Shaun~A Forth}.}
  \bibinfo{year}{2006}\natexlab{}.
\newblock \showarticletitle{An efficient overloaded implementation of forward
  mode automatic differentiation in MATLAB}.
\newblock \bibinfo{journal}{\emph{ACM Transactions on Mathematical Software
  (TOMS)}} \bibinfo{volume}{32}, \bibinfo{number}{2} (\bibinfo{year}{2006}),
  \bibinfo{pages}{195--222}.
\newblock


\bibitem[\protect\citeauthoryear{Gill, Launchbury, and Peyton~Jones}{Gill
  et~al\mbox{.}}{1993}]%
        {foldr-fusion-1}
\bibfield{author}{\bibinfo{person}{Andrew Gill}, \bibinfo{person}{John
  Launchbury}, {and} \bibinfo{person}{Simon~L Peyton~Jones}.}
  \bibinfo{year}{1993}\natexlab{}.
\newblock \showarticletitle{A short cut to deforestation}
  \emph{(\bibinfo{series}{FPCA})}. ACM, \bibinfo{pages}{223--232}.
\newblock


\bibitem[\protect\citeauthoryear{Grelck and Scholz}{Grelck and Scholz}{2006}]%
        {Grelck2006}
\bibfield{author}{\bibinfo{person}{Clemens Grelck} {and}
  \bibinfo{person}{Sven-Bodo Scholz}.} \bibinfo{year}{2006}\natexlab{}.
\newblock \showarticletitle{{SAC}---{A} Functional Array Language for Efficient
  Multi-threaded Execution}.
\newblock \bibinfo{journal}{\emph{Int. Journal of Parallel Programming}}
  \bibinfo{volume}{34}, \bibinfo{number}{4} (\bibinfo{year}{2006}),
  \bibinfo{pages}{383--427}.
\newblock
\showISSN{1573-7640}


\bibitem[\protect\citeauthoryear{Hascoet and Pascual}{Hascoet and
  Pascual}{2013}]%
        {tapenade}
\bibfield{author}{\bibinfo{person}{Laurent Hascoet} {and}
  \bibinfo{person}{Val{\'e}rie Pascual}.} \bibinfo{year}{2013}\natexlab{}.
\newblock \showarticletitle{The Tapenade Automatic Differentiation Tool:
  Principles, Model, and Specification}.
\newblock \bibinfo{journal}{\emph{ACM Trans. Math. Softw.}}
  \bibinfo{volume}{39}, \bibinfo{number}{3}, Article \bibinfo{articleno}{20}
  (\bibinfo{date}{May} \bibinfo{year}{2013}), \bibinfo{numpages}{43}~pages.
\newblock
\showISSN{0098-3500}
\urldef\tempurl%
\url{https://doi.org/10.1145/2450153.2450158}
\showDOI{\tempurl}


\bibitem[\protect\citeauthoryear{Henriksen, Serup, Elsman, Henglein, and
  Oancea}{Henriksen et~al\mbox{.}}{2017}]%
        {henriksen2017futhark}
\bibfield{author}{\bibinfo{person}{Troels Henriksen}, \bibinfo{person}{Niels~GW
  Serup}, \bibinfo{person}{Martin Elsman}, \bibinfo{person}{Fritz Henglein},
  {and} \bibinfo{person}{Cosmin~E Oancea}.} \bibinfo{year}{2017}\natexlab{}.
\newblock \showarticletitle{Futhark: purely functional GPU-programming with
  nested parallelism and in-place array updates}. In
  \bibinfo{booktitle}{\emph{Proceedings of the 38th ACM SIGPLAN Conference on
  Programming Language Design and Implementation}}. ACM,
  \bibinfo{pages}{556--571}.
\newblock


\bibitem[\protect\citeauthoryear{Hivert and Thi{\'e}ry}{Hivert and
  Thi{\'e}ry}{2004}]%
        {hivert2004mupad}
\bibfield{author}{\bibinfo{person}{Florent Hivert} {and} \bibinfo{person}{N
  Thi{\'e}ry}.} \bibinfo{year}{2004}\natexlab{}.
\newblock \showarticletitle{MuPAD-Combinat, an open-source package for research
  in algebraic combinatorics}.
\newblock \bibinfo{journal}{\emph{S{\'e}m. Lothar. Combin}}
  \bibinfo{volume}{51} (\bibinfo{year}{2004}), \bibinfo{pages}{70}.
\newblock


\bibitem[\protect\citeauthoryear{Hogan}{Hogan}{2014}]%
        {adept}
\bibfield{author}{\bibinfo{person}{Robin~J. Hogan}.}
  \bibinfo{year}{2014}\natexlab{}.
\newblock \showarticletitle{Fast Reverse-Mode Automatic Differentiation Using
  Expression Templates in C++}.
\newblock \bibinfo{journal}{\emph{ACM Trans. Math. Softw.}}
  \bibinfo{volume}{40}, \bibinfo{number}{4}, Article \bibinfo{articleno}{26}
  (\bibinfo{date}{July} \bibinfo{year}{2014}), \bibinfo{numpages}{16}~pages.
\newblock
\showISSN{0098-3500}
\urldef\tempurl%
\url{https://doi.org/10.1145/2560359}
\showDOI{\tempurl}


\bibitem[\protect\citeauthoryear{Huang and Chau}{Huang and Chau}{2008}]%
        {huang2008new}
\bibfield{author}{\bibinfo{person}{Zhi-Kai Huang} {and}
  \bibinfo{person}{Kwok-Wing Chau}.} \bibinfo{year}{2008}\natexlab{}.
\newblock \showarticletitle{A new image thresholding method based on Gaussian
  mixture model}.
\newblock \bibinfo{journal}{\emph{Appl. Math. Comput.}} \bibinfo{volume}{205},
  \bibinfo{number}{2} (\bibinfo{year}{2008}), \bibinfo{pages}{899--907}.
\newblock


\bibitem[\protect\citeauthoryear{Hudak}{Hudak}{1996}]%
        {hudak-dsl}
\bibfield{author}{\bibinfo{person}{Paul Hudak}.}
  \bibinfo{year}{1996}\natexlab{}.
\newblock \showarticletitle{Building Domain-specific Embedded Languages}.
\newblock \bibinfo{journal}{\emph{ACM Comput. Surv.}} \bibinfo{volume}{28},
  \bibinfo{number}{4} (\bibinfo{date}{Dec.} \bibinfo{year}{1996}).
\newblock
\showISSN{0360-0300}


\bibitem[\protect\citeauthoryear{Iverson}{Iverson}{1962}]%
        {iverson1962programming}
\bibfield{author}{\bibinfo{person}{Kenneth~E Iverson}.}
  \bibinfo{year}{1962}\natexlab{}.
\newblock \showarticletitle{{A Programming Language}}. In
  \bibinfo{booktitle}{\emph{Proceedings of the May 1-3, 1962, spring joint
  computer conference}}. ACM, \bibinfo{pages}{345--351}.
\newblock


\bibitem[\protect\citeauthoryear{Ji-rong Wen}{Ji-rong Wen}{2002}]%
        {2002:QCU:503104.503108}
\bibfield{author}{\bibinfo{person}{Hong-Jiang~Zhang Ji-rong Wen, Jian-Yun
  Nie~and}.} \bibinfo{year}{2002}\natexlab{}.
\newblock \showarticletitle{Query Clustering Using User Logs}.
\newblock \bibinfo{journal}{\emph{ACM Transactions on Information Systems}}
  \bibinfo{volume}{20}, \bibinfo{number}{1} (\bibinfo{year}{2002}).
\newblock


\bibitem[\protect\citeauthoryear{Karczmarczuk}{Karczmarczuk}{1999}]%
        {karczmarczuk1999functional}
\bibfield{author}{\bibinfo{person}{Jerzy Karczmarczuk}.}
  \bibinfo{year}{1999}\natexlab{}.
\newblock \showarticletitle{Functional differentiation of computer programs}.
\newblock \bibinfo{journal}{\emph{ACM SIGPLAN Notices}} \bibinfo{volume}{34},
  \bibinfo{number}{1} (\bibinfo{year}{1999}), \bibinfo{pages}{195--203}.
\newblock


\bibitem[\protect\citeauthoryear{Khan and Barton}{Khan and Barton}{2015}]%
        {khan2015vector}
\bibfield{author}{\bibinfo{person}{Kamil~A Khan} {and} \bibinfo{person}{Paul~I
  Barton}.} \bibinfo{year}{2015}\natexlab{}.
\newblock \showarticletitle{A vector forward mode of automatic differentiation
  for generalized derivative evaluation}.
\newblock \bibinfo{journal}{\emph{Optimization Methods and Software}}
  \bibinfo{volume}{30}, \bibinfo{number}{6} (\bibinfo{year}{2015}),
  \bibinfo{pages}{1185--1212}.
\newblock


\bibitem[\protect\citeauthoryear{Levenberg}{Levenberg}{1944}]%
        {levenberg1944method}
\bibfield{author}{\bibinfo{person}{Kenneth Levenberg}.}
  \bibinfo{year}{1944}\natexlab{}.
\newblock \showarticletitle{A method for the solution of certain non-linear
  problems in least squares}.
\newblock \bibinfo{journal}{\emph{Quarterly of applied mathematics}}
  \bibinfo{volume}{2}, \bibinfo{number}{2} (\bibinfo{year}{1944}),
  \bibinfo{pages}{164--168}.
\newblock


\bibitem[\protect\citeauthoryear{Liu, Yang, Fan, He, and Wang}{Liu
  et~al\mbox{.}}{2010}]%
        {liu2010distributed}
\bibfield{author}{\bibinfo{person}{Chao Liu}, \bibinfo{person}{Hung-chih Yang},
  \bibinfo{person}{Jinliang Fan}, \bibinfo{person}{Li-Wei He}, {and}
  \bibinfo{person}{Yi-Min Wang}.} \bibinfo{year}{2010}\natexlab{}.
\newblock \showarticletitle{Distributed nonnegative matrix factorization for
  web-scale dyadic data analysis on mapreduce}. In
  \bibinfo{booktitle}{\emph{Proceedings of the 19th international conference on
  World wide web}}. ACM, \bibinfo{pages}{681--690}.
\newblock


\bibitem[\protect\citeauthoryear{Maclaurin, Duvenaud, and Adams}{Maclaurin
  et~al\mbox{.}}{2015}]%
        {maclaurin2015autograd}
\bibfield{author}{\bibinfo{person}{Dougal Maclaurin}, \bibinfo{person}{David
  Duvenaud}, {and} \bibinfo{person}{Ryan~P Adams}.}
  \bibinfo{year}{2015}\natexlab{}.
\newblock \showarticletitle{Autograd: Effortless gradients in numpy}. In
  \bibinfo{booktitle}{\emph{ICML 2015 AutoML Workshop}}.
\newblock


\bibitem[\protect\citeauthoryear{Marquardt}{Marquardt}{1963}]%
        {marquardt1963algorithm}
\bibfield{author}{\bibinfo{person}{Donald~W Marquardt}.}
  \bibinfo{year}{1963}\natexlab{}.
\newblock \showarticletitle{An algorithm for least-squares estimation of
  nonlinear parameters}.
\newblock \bibinfo{journal}{\emph{Journal of the society for Industrial and
  Applied Mathematics}} \bibinfo{volume}{11}, \bibinfo{number}{2}
  (\bibinfo{year}{1963}), \bibinfo{pages}{431--441}.
\newblock


\bibitem[\protect\citeauthoryear{Mor{\'e}}{Mor{\'e}}{1978}]%
        {more1978levenberg}
\bibfield{author}{\bibinfo{person}{Jorge~J Mor{\'e}}.}
  \bibinfo{year}{1978}\natexlab{}.
\newblock \showarticletitle{The Levenberg-Marquardt algorithm: implementation
  and theory}.
\newblock In \bibinfo{booktitle}{\emph{Numerical analysis}}.
  \bibinfo{publisher}{Springer}, \bibinfo{pages}{105--116}.
\newblock


\bibitem[\protect\citeauthoryear{Narayanan, Norris, and Winnicka}{Narayanan
  et~al\mbox{.}}{2010}]%
        {narayanan2010adic2}
\bibfield{author}{\bibinfo{person}{Sri Hari~Krishna Narayanan},
  \bibinfo{person}{Boyana Norris}, {and} \bibinfo{person}{Beata Winnicka}.}
  \bibinfo{year}{2010}\natexlab{}.
\newblock \showarticletitle{ADIC2: Development of a component source
  transformation system for differentiating C and C++}.
\newblock \bibinfo{journal}{\emph{Procedia Computer Science}}
  \bibinfo{volume}{1}, \bibinfo{number}{1} (\bibinfo{year}{2010}),
  \bibinfo{pages}{1845--1853}.
\newblock


\bibitem[\protect\citeauthoryear{Nielsen and Sun}{Nielsen and Sun}{2016}]%
        {nielsen2016guaranteed}
\bibfield{author}{\bibinfo{person}{Frank Nielsen} {and} \bibinfo{person}{Ke
  Sun}.} \bibinfo{year}{2016}\natexlab{}.
\newblock \showarticletitle{Guaranteed bounds on information-theoretic measures
  of univariate mixtures using piecewise log-sum-exp inequalities}.
\newblock \bibinfo{journal}{\emph{Entropy}} \bibinfo{volume}{18},
  \bibinfo{number}{12} (\bibinfo{year}{2016}), \bibinfo{pages}{442}.
\newblock


\bibitem[\protect\citeauthoryear{Norvig}{Norvig}{1992}]%
        {Norvig92}
\bibfield{author}{\bibinfo{person}{Peter Norvig}.}
  \bibinfo{year}{1992}\natexlab{}.
\newblock \bibinfo{booktitle}{\emph{Paradigms of Artificial Intelligence
  Programming: Case Studies in Common Lisp}}.
\newblock \bibinfo{publisher}{Morgan Kaufmann}.
\newblock


\bibitem[\protect\citeauthoryear{Pearlmutter and Siskind}{Pearlmutter and
  Siskind}{2007}]%
        {pearlmutter2007lazy}
\bibfield{author}{\bibinfo{person}{Barak~A Pearlmutter} {and}
  \bibinfo{person}{Jeffrey~Mark Siskind}.} \bibinfo{year}{2007}\natexlab{}.
\newblock \showarticletitle{Lazy multivariate higher-order forward-mode AD}. In
  \bibinfo{booktitle}{\emph{ACM SIGPLAN Notices}}, Vol.~\bibinfo{volume}{42}.
  ACM, \bibinfo{pages}{155--160}.
\newblock


\bibitem[\protect\citeauthoryear{Pearlmutter and Siskind}{Pearlmutter and
  Siskind}{2008}]%
        {pearlmutter2008reverse}
\bibfield{author}{\bibinfo{person}{Barak~A Pearlmutter} {and}
  \bibinfo{person}{Jeffrey~Mark Siskind}.} \bibinfo{year}{2008}\natexlab{}.
\newblock \showarticletitle{Reverse-mode AD in a functional framework: Lambda
  the ultimate backpropagator}.
\newblock \bibinfo{journal}{\emph{ACM Transactions on Programming Languages and
  Systems (TOPLAS)}} \bibinfo{volume}{30}, \bibinfo{number}{2}
  (\bibinfo{year}{2008}), \bibinfo{pages}{7}.
\newblock


\bibitem[\protect\citeauthoryear{Puschel, Moura, Johnson, Padua, Veloso,
  Singer, Xiong, Franchetti, Gacic, Voronenko, et~al\mbox{.}}{Puschel
  et~al\mbox{.}}{2005}]%
        {spiral}
\bibfield{author}{\bibinfo{person}{Markus Puschel},
  \bibinfo{person}{Jos{\'e}~MF Moura}, \bibinfo{person}{Jeremy~R Johnson},
  \bibinfo{person}{David Padua}, \bibinfo{person}{Manuela~M Veloso},
  \bibinfo{person}{Bryan~W Singer}, \bibinfo{person}{Jianxin Xiong},
  \bibinfo{person}{Franz Franchetti}, \bibinfo{person}{Aca Gacic},
  \bibinfo{person}{Yevgen Voronenko}, {et~al\mbox{.}}}
  \bibinfo{year}{2005}\natexlab{}.
\newblock \showarticletitle{{SPIRAL:} Code generation for {DSP} transforms}.
\newblock \bibinfo{journal}{\emph{Proc. IEEE}} \bibinfo{volume}{93},
  \bibinfo{number}{2} (\bibinfo{year}{2005}), \bibinfo{pages}{232--275}.
\newblock


\bibitem[\protect\citeauthoryear{Ragan-Kelley, Barnes, Adams, Paris, Durand,
  and Amarasinghe}{Ragan-Kelley et~al\mbox{.}}{2013}]%
        {ragan2013halide}
\bibfield{author}{\bibinfo{person}{Jonathan Ragan-Kelley},
  \bibinfo{person}{Connelly Barnes}, \bibinfo{person}{Andrew Adams},
  \bibinfo{person}{Sylvain Paris}, \bibinfo{person}{Fr{\'e}do Durand}, {and}
  \bibinfo{person}{Saman Amarasinghe}.} \bibinfo{year}{2013}\natexlab{}.
\newblock \showarticletitle{Halide: A Language and Compiler for Optimizing
  Parallelism, Locality, and Recomputation in Image Processing Pipelines}
  \emph{(\bibinfo{series}{PLDI '13})}.
\newblock


\bibitem[\protect\citeauthoryear{Revels, Lubin, and Papamarkou}{Revels
  et~al\mbox{.}}{2016}]%
        {revels2016forward}
\bibfield{author}{\bibinfo{person}{Jarrett Revels}, \bibinfo{person}{Miles
  Lubin}, {and} \bibinfo{person}{Theodore Papamarkou}.}
  \bibinfo{year}{2016}\natexlab{}.
\newblock \showarticletitle{Forward-mode automatic differentiation in Julia}.
\newblock \bibinfo{journal}{\emph{arXiv preprint arXiv:1607.07892}}
  (\bibinfo{year}{2016}).
\newblock


\bibitem[\protect\citeauthoryear{Shaikhha, Fitzgibbon, Peyton~Jones, and
  Vytiniotis}{Shaikhha et~al\mbox{.}}{2017}]%
        {dps_fhpc}
\bibfield{author}{\bibinfo{person}{Amir Shaikhha}, \bibinfo{person}{Andrew
  Fitzgibbon}, \bibinfo{person}{Simon Peyton~Jones}, {and}
  \bibinfo{person}{Dimitrios Vytiniotis}.} \bibinfo{year}{2017}\natexlab{}.
\newblock \showarticletitle{Destination-passing Style for Efficient Memory
  Management}. In \bibinfo{booktitle}{\emph{Proceedings of the 6th ACM SIGPLAN
  International Workshop on Functional High-Performance Computing}}
  \emph{(\bibinfo{series}{FHPC 2017})}. \bibinfo{publisher}{ACM},
  \bibinfo{address}{New York, NY, USA}, \bibinfo{pages}{12--23}.
\newblock
\showISBNx{978-1-4503-5181-2}
\urldef\tempurl%
\url{https://doi.org/10.1145/3122948.3122949}
\showDOI{\tempurl}


\bibitem[\protect\citeauthoryear{Siskind and Pearlmutter}{Siskind and
  Pearlmutter}{2005}]%
        {siskind2005perturbation}
\bibfield{author}{\bibinfo{person}{Jeffrey~Mark Siskind} {and}
  \bibinfo{person}{Barak~A Pearlmutter}.} \bibinfo{year}{2005}\natexlab{}.
\newblock \showarticletitle{Perturbation confusion and referential
  transparency: Correct functional implementation of forward-mode AD}.
\newblock  (\bibinfo{year}{2005}).
\newblock


\bibitem[\protect\citeauthoryear{Spampinato and P{\"u}schel}{Spampinato and
  P{\"u}schel}{[n. d.]}]%
        {spampinato2016basic}
\bibfield{author}{\bibinfo{person}{Daniele~G Spampinato} {and}
  \bibinfo{person}{Markus P{\"u}schel}.} \bibinfo{year}{[n. d.]}\natexlab{}.
\newblock \showarticletitle{A basic linear algebra compiler for structured
  matrices}. In \bibinfo{booktitle}{\emph{CGO '16}}. ACM.
\newblock


\bibitem[\protect\citeauthoryear{Sra and Dhillon}{Sra and Dhillon}{2006}]%
        {Sra06nonnegativematrix}
\bibfield{author}{\bibinfo{person}{Suvrit Sra} {and}
  \bibinfo{person}{Inderjit~S. Dhillon}.} \bibinfo{year}{2006}\natexlab{}.
\newblock \bibinfo{booktitle}{\emph{Nonnegative matrix approximation:
  algorithms and applications}}.
\newblock \bibinfo{type}{{T}echnical {R}eport}.
\newblock


\bibitem[\protect\citeauthoryear{Srajer, Kukelova, and Fitzgibbon}{Srajer
  et~al\mbox{.}}{2016}]%
        {srajerbenchmark}
\bibfield{author}{\bibinfo{person}{Filip Srajer}, \bibinfo{person}{Zuzana
  Kukelova}, {and} \bibinfo{person}{Andrew Fitzgibbon}.}
  \bibinfo{year}{2016}\natexlab{}.
\newblock \showarticletitle{A Benchmark of Selected Algorithmic Differentiation
  Tools on Some Problems in Machine Learning and Computer Vision}.
\newblock  (\bibinfo{year}{2016}).
\newblock


\bibitem[\protect\citeauthoryear{Steuwer, Fensch, Lindley, and Dubach}{Steuwer
  et~al\mbox{.}}{2015}]%
        {Steuwer:2015:GPP:2784731.2784754}
\bibfield{author}{\bibinfo{person}{Michel Steuwer}, \bibinfo{person}{Christian
  Fensch}, \bibinfo{person}{Sam Lindley}, {and} \bibinfo{person}{Christophe
  Dubach}.} \bibinfo{year}{2015}\natexlab{}.
\newblock \showarticletitle{Generating Performance Portable Code Using Rewrite
  Rules: From High-level Functional Expressions to High-performance OpenCL
  Code}. In \bibinfo{booktitle}{\emph{Proceedings of the 20th ACM SIGPLAN
  International Conference on Functional Programming}}
  \emph{(\bibinfo{series}{ICFP 2015})}. \bibinfo{publisher}{ACM},
  \bibinfo{address}{New York, NY, USA}, \bibinfo{pages}{205--217}.
\newblock
\showISBNx{978-1-4503-3669-7}
\urldef\tempurl%
\url{https://doi.org/10.1145/2784731.2784754}
\showDOI{\tempurl}


\bibitem[\protect\citeauthoryear{Sujeeth, Lee, Brown, Rompf, Chafi, Wu, Atreya,
  Odersky, and Olukotun}{Sujeeth et~al\mbox{.}}{2011}]%
        {sujeeth2011optiml}
\bibfield{author}{\bibinfo{person}{Arvind Sujeeth}, \bibinfo{person}{HyoukJoong
  Lee}, \bibinfo{person}{Kevin Brown}, \bibinfo{person}{Tiark Rompf},
  \bibinfo{person}{Hassan Chafi}, \bibinfo{person}{Michael Wu},
  \bibinfo{person}{Anand Atreya}, \bibinfo{person}{Martin Odersky}, {and}
  \bibinfo{person}{Kunle Olukotun}.} \bibinfo{year}{2011}\natexlab{}.
\newblock \showarticletitle{{OptiML: An Implicitly Parallel Domain-Specific
  Language for Machine Learning}} \emph{(\bibinfo{series}{ICML '11})}.
  \bibinfo{pages}{609--616}.
\newblock


\bibitem[\protect\citeauthoryear{Svenningsson}{Svenningsson}{2002}]%
        {Svenningsson:2002:SFA:581478.581491}
\bibfield{author}{\bibinfo{person}{Josef Svenningsson}.}
  \bibinfo{year}{2002}\natexlab{}.
\newblock \showarticletitle{Shortcut Fusion for Accumulating Parameters \&
  Zip-like Functions} \emph{(\bibinfo{series}{ICFP '02})}.
  \bibinfo{publisher}{ACM}, \bibinfo{pages}{124--132}.
\newblock
\showISBNx{1-58113-487-8}


\bibitem[\protect\citeauthoryear{Svensson and Svenningsson}{Svensson and
  Svenningsson}{2014}]%
        {Svensson:2014:DPA:2636228.2636231}
\bibfield{author}{\bibinfo{person}{Bo~Joel Svensson} {and}
  \bibinfo{person}{Josef Svenningsson}.} \bibinfo{year}{2014}\natexlab{}.
\newblock \showarticletitle{Defunctionalizing Push Arrays}
  \emph{(\bibinfo{series}{FHPC '14})}. \bibinfo{publisher}{ACM},
  \bibinfo{address}{NY, USA}, \bibinfo{pages}{43--52}.
\newblock
\showISBNx{978-1-4503-3040-4}


\bibitem[\protect\citeauthoryear{Triggs, McLauchlan, Hartley, and
  Fitzgibbon}{Triggs et~al\mbox{.}}{1999}]%
        {triggs1999bundle}
\bibfield{author}{\bibinfo{person}{Bill Triggs}, \bibinfo{person}{Philip~F
  McLauchlan}, \bibinfo{person}{Richard~I Hartley}, {and}
  \bibinfo{person}{Andrew~W Fitzgibbon}.} \bibinfo{year}{1999}\natexlab{}.
\newblock \showarticletitle{Bundle adjustment—a modern synthesis}. In
  \bibinfo{booktitle}{\emph{Inter. workshop on vision algorithms}}. Springer,
  \bibinfo{pages}{298--372}.
\newblock


\bibitem[\protect\citeauthoryear{Wadler}{Wadler}{1988}]%
        {deforestation}
\bibfield{author}{\bibinfo{person}{Philip Wadler}.}
  \bibinfo{year}{1988}\natexlab{}.
\newblock \showarticletitle{Deforestation: Transforming programs to eliminate
  trees}. In \bibinfo{booktitle}{\emph{ESOP'88}}. Springer,
  \bibinfo{pages}{344--358}.
\newblock


\bibitem[\protect\citeauthoryear{Weinstein and Rao}{Weinstein and Rao}{2016}]%
        {weinstein2016algorithm}
\bibfield{author}{\bibinfo{person}{Matthew~J Weinstein} {and}
  \bibinfo{person}{Anil~V Rao}.} \bibinfo{year}{2016}\natexlab{}.
\newblock \showarticletitle{Algorithm: ADiGator, a toolbox for the algorithmic
  differentiation of mathematical functions in MATLAB using source
  transformation via operator overloading}.
\newblock \bibinfo{journal}{\emph{ACM Trans. Math. Softw}}
  (\bibinfo{year}{2016}).
\newblock


\bibitem[\protect\citeauthoryear{Zach}{Zach}{2014}]%
        {zach2014robust}
\bibfield{author}{\bibinfo{person}{Christopher Zach}.}
  \bibinfo{year}{2014}\natexlab{}.
\newblock \showarticletitle{Robust bundle adjustment revisited}. In
  \bibinfo{booktitle}{\emph{European Conference on Computer Vision}}. Springer,
  \bibinfo{pages}{772--787}.
\newblock


\end{thebibliography}
